\documentclass[twocolumn,useAMS,usenatbib]{mn2e}
\usepackage{natbib}
\usepackage{amsmath,latexsym,amssymb}
\usepackage{epsfig,graphicx}
\usepackage{float}
\usepackage{multicol}
\usepackage{color}

\DeclareGraphicsExtensions{.pdf,.png,.jpg,.eps}
\newcommand{\wgg}{\ensuremath{w_{gg}}}
\newcommand{\wgp}{\ensuremath{w_{g+}}}
\newcommand{\wgx}{\ensuremath{w_{g\times}}}
\newcommand{\wpp}{\ensuremath{w_{++}}}
\newcommand{\wxx}{\ensuremath{w_{\times\times}}}
\newcommand{\mpch}{\ensuremath{h^{-1}\text{Mpc}}}

\newcommand{\ia}{intrinsic alignments}
\newcommand{\IA}{Intrinsic alignments}

\newcommand{\apj}{ApJ}%                                         % Journal abbreviations                                                                                                   
\newcommand{\apjs}{ApJS}

\newcommand{\apjl}{ApJL}
\newcommand{\aap}{A{\&}A}
\newcommand{\aaps}{A{\&}AS}
\newcommand{\mnras}{MNRAS}
\newcommand{\aj}{AJ}
\newcommand{\araa}{ARAA}
\newcommand{\pasp}{PASP}
\def\reff@jnl#1{{\rm#1\/}}

\def\aj{\reff@jnl{AJ}}                  % Astronomical Journal
\def\araa{\reff@jnl{ARA\&A}}            % Annual Review of Astron and Astrophys
\def\apj{\reff@jnl{ApJ}}                        % Astrophysical Journal
\def\apjl{\reff@jnl{ApJ}}               % Astrophysical Journal, Letters
\def\apjs{\reff@jnl{ApJS}}              % Astrophysical Journal, Supplement
\def\apss{\reff@jnl{Ap\&SS}}            % Astrophysics and Space Science
\def\aap{\reff@jnl{A\&A}}               % Astronomy and Astrophysics
\def\aapr{\reff@jnl{A\&A~Rev.}}         % Astronomy and Astrophysics Reviews
\def\aaps{\reff@jnl{A\&AS}}             % Astronomy and Astrophysics, Supplement
\def\baas{\reff@jnl{BAAS}}              % Bulletin of the AAS
\def\jrasc{\reff@jnl{JRASC}}            % Journal of the RAS of Canada
\def\memras{\reff@jnl{MmRAS}}           % Memoirs of the RAS
\def\mnras{\reff@jnl{MNRAS}}            % Monthly Notices of the RAS
\def\physrep{\reff@jnl{Phys.Rep.}}
\def\pra{\reff@jnl{Phys.Rev.A}}         % Physical Review A: General Physics
\def\prb{\reff@jnl{Phys.Rev.B}}         % Physical Review B: Solid State
\def\prc{\reff@jnl{Phys.Rev.C}}         % Physical Review C
\def\prd{\reff@jnl{Phys.Rev.D}}         % Physical Review D
\def\prl{\reff@jnl{Phys.Rev.Lett}}      % Physical Review Letters
\def\pasp{\reff@jnl{PASP}}              % Publications of the ASP
\def\pasj{\reff@jnl{PASJ}}              % Publications of the ASJ
\def\skytel{\reff@jnl{S\&T}}            % Sky and Telescope
\def\solphys{\reff@jnl{Solar~Phys.}}    % Solar Physics
\def\sovast{\reff@jnl{Soviet~Ast.}}     % Soviet Astronomy
\def\ssr{\reff@jnl{Space~Sci.Rev.}}     % Space Science Reviews
\def\nat{\reff@jnl{Nature}}             % Nature

\newcommand{\hmpc}{\ensuremath{h^{-1}\mathrm{Mpc}}}

\newcommand{\Msun}{M_{\odot}}

\newcommand{\beq}{\begin{equation}}
\newcommand{\eeq}{\end{equation}}
\newcommand{\beqa}{\begin{eqnarray}}
\newcommand{\eeqa}{\end{eqnarray}}

\graphicspath{{figures/dr11/}}

\newcommand{\rmD}{{\rm D}}
\newcommand{\rmS}{{\rm S}}

\newcommand{\vc}[1]{{\vec{#1}}}
\newcommand{\greal}{{\gamma_{\rm I}^{\rm real}}}
\newcommand{\tgreal}{{\widetilde\gamma_{\rm I}^{\rm real}}}
\newcommand{\gred}{{\gamma_{\rm I}^{\rm red}}}

\usepackage{bm}
\renewcommand{\vec}[1]{\boldsymbol{\mathbf{#1}}}

\title[Intrinsic Alignments in BOSS]{Intrinsic alignments of SDSS-III BOSS LOWZ sample galaxies}

\author[Singh et~al.]{
	Sukhdeep Singh$^1$\thanks{\tt sukhdeep@cmu.edu}, 
	Rachel Mandelbaum$^{1}$,
    Surhud More$^{2}$
\\
	 $^1$McWilliams Center for Cosmology, Department of Physics, Carnegie Mellon University, Pittsburgh, PA 	
	 15213, USA\\
     $^2$Kavli Institute for the Physics and Mathematics of the Universe (WPI), TODIAS, The University of Tokyo, 
Chiba, 277-8583, Japan \\
}
\date{\today}

\begin{document}
	\maketitle
	\begin{abstract}
		 Intrinsic alignments (IA) of galaxies, i.e. correlations of galaxy shapes with each other (II) or with the density field (gI), are {potentially} a major astrophysical source of contamination for weak lensing surveys. 
		 We present the results of IA measurements of galaxies on $0.1$--$200h^{-1}$Mpc scales using the SDSS-III BOSS LOWZ sample, in the redshift range $0.16<z<0.36$. We  
		 extend the existing IA measurements for spectroscopic LRGs to lower luminosities, and show that the luminosity dependence of large-scale IA can be well-described by a power law. Within the limited redshift and color 
		 range of our sample, we observe no significant redshift or color 
		 dependence of IA. We measure the halo mass of galaxies using galaxy-galaxy lensing, and show that the mass dependence of large-scale IA is also well described by a power law. We detect variations in  the 
		 scale dependence of IA with mass and luminosity, which underscores the need to use flexible templates in order to remove the IA signal. We also study the 
		 environment 
		 dependence of IA by splitting the sample into field and group galaxies, which are further split into satellite and central galaxies. We show that group central galaxies are aligned with their halos at small scales and also 
		 are aligned with the tidal fields out to large scales. {We also detect the radial alignments of satellite galaxies within groups}.  These results 
		 can be used to construct better intrinsic alignment models for removal of this contaminant
         to the weak lensing signal. 
	\end{abstract}
	
	\begin{keywords}
		 galaxies: evolution\ -- cosmology: observations --large-scale structure of Universe\ --gravitational lensing: weak
	\end{keywords}

	\section{Introduction}\label{sec:intro}

		The deflection of light rays due to the gravitational effects of matter, gravitational lensing, changes both the observed shape and size of distant galaxies. Weak lensing, the statistical study of such tiny shape distortions, has 
		emerged as an important tool in cosmology to study and map the distribution of dark matter as well as to study the effects 
		of dark energy. Due to its sensitivity to the total matter content, weak lensing is theoretically a clean probe for dark matter and dark energy, but it is affected by a number of potential systematic sources 
		\citep[see, e.g.,][]{Weinberg2013}. Much effort has been put into controlling the observational systematics arising from the effect of the point-spread function (PSF) on the estimated galaxy shape distortions. Future 
		surveys like LSST\footnote{\texttt{http://www.lsst.org/}} \citep{2009arXiv0912.0201L}, Euclid\footnote{\texttt{http://sci.esa.int/euclid/}, \texttt{http://www.euclid-ec.org}} \citep{2011arXiv1110.3193L} and WFIRST-AFTA
		\footnote{\texttt{http://wfirst.gsfc.nasa.gov/}} \citep{2013arXiv1305.5422S} aim to bring these systematics below the percent level, with comparable improvement being needed on 
		astrophysical sources of contamination to the weak lensing measurements.
		
		\IA{} of galaxies is perhaps the most important astrophysical systematic for weak lensing {\citep[for a review see,][]{Troxel2014}}. 
		It is the coherent alignments of the shapes of physically nearby galaxies which can mimic the weak lensing signal. One likely explanation for \ia{} is that they are an environmental effect, whereby the local or large-scale tidal 
		fields can shear and align the galaxy shapes, 
		producing shape correlations. These correlations violate the assumption in weak lensing studies that observed correlations in galaxy shapes are only caused by gravitational lensing. \cite{Hirata2004} developed the 
		linear alignment model for the \ia{}, assuming that \ia{} are set by tidal fields at the time of galaxy formation \citep{Catelan2001}. This model was later extended by \cite{Bridle2007} to non-linear scales using the full 
		non-linear matter power spectrum. To extend this model inside of dark matter halos, \cite{Schneider2010} developed the halo model of \ia{}, which assumes that satellites are aligned radially within the halo, with central 
		galaxies (BCGs or BGGs) being preferentially aligned with the shape of the halo. {There is also extensive literature on spin alignments of spiral galaxies which we will not explore in this paper \citep[see, e.g.,][]{Crittenden2001,Capranico2013,Zhang2015}.} Constraining these models observationally is important for all  techniques for mitigating \ia{} in future weak lensing surveys that involve 
		marginalizing over models with some free parameters \citep[e.g.,][]{2010A&A...523A...1J}; these techniques require valid models, and priors on the model parameters.
		
		\IA{} are also interesting from the galaxy formation perspective, as the alignments of galaxies with their dark matter halos and the alignments of dark matter halos themselves are intricately related to the processes 
		involved in galaxy formation and evolution. {Many 
        studies have concentrated on studying \ia{} of dark matter halos in cosmological simulations 
		\citep[for
        example,][]{Croft2000,Hopkins2005,Basilakos2006,Heymans2006,Kang2007,2007ApJ...671.1135K,Faltenbacher2008,Knebe2008,Pereira2008,Schneider2012,Forero-Romero2014}. 
		These studies have found that dark matter halo shapes tend to align with each other and with
        the large scale structure \citep[][]{Schneider2012,Forero-Romero2014}. These alignments
        depend on the halo properties such as mass, radii and redshift. More massive halos in general have higher ellipticities and show stronger alignments. Ellipticities and alignments also increase with the redshift, with higher redshift halos being more elliptical and showing stronger alignments \citep[see, for example,][]{Hopkins2005,Figueroa2010,Wang2011,Schneider2012}. This is consistent with the picture of alignments being imprinted at the time of formation which can then get disrupted by various processes such as halo mergers and hence alignments decrease with redshift. There is also radial dependence, where the ellipticity and the alignment signal varies with the radii used to measure the shapes of the halos, with outer radii being more elliptical and showing stronger alignments \citep[see, for example,][]{Hopkins2005,Schneider2012,Tenneti2014,Tenneti2014b}. }

{  At small scales, the subhalos
    within larger halos tend to align radially within their
        host halos \citep{Kuhlen2007,Knebe2008,Pereira2008,Faltenbacher2008}. These radial
        alignments primarily originate from tidal torquing of galaxies within host halo
        \citep{Pereira2008}, are independent of host halo mass \citep{Knebe2008,Pereira2008}, depend
        on the distance of the subhalo from the center of the host halo, and can be observed out to
        large distances of $\sim 6 R_{\mathrm{vir}}$ \citep{Faltenbacher2008}.  The satellite
        alignment signal also has some contribution from the anisotropic satellite distribution,
        with satellites distributed preferentially along the host halo major axis \citep{Faltenbacher2008}.}
        
        { Several studies have also looked at the relation between dark matter shapes and galaxy shapes by populating the dark matter halos with galaxies using semi analytic models \citep{Okumura2009,Faltenbacher2008,Schneider2012,Joachimi2013I,Joachimi2013II}. For example,
		\cite{Joachimi2013I,Joachimi2013II} studied shapes and \ia{} of COSMOS galaxies by populating dark 
		matter halos in the Millennium simulation \citep{2005Natur.435..629S} with galaxies using semi-analytical methods, finding a strong dependence of \ia{} on scale, redshift and galaxy type. \cite{Faltenbacher2008} also compared the results from semi-analytic modeling and SDSS DR6 LRGs and found that a mean misalignment angel of $\sim 25$ degrees between the dark matter shape and galaxy shapes is necessary to explain the observations \citep[see also][]{Okumura2009,Schneider2012}.
		\cite{Tenneti2014} also studied the orientations of stellar components of galaxies 
		within their halos using the MassiveBlack-II (MB-II) hydrodynamical simulation \citep{Khandai2014}, 
		 and found that galaxy shapes are preferentially aligned with their dark matter halos with
         an 
		average misalignment angle of $\sim 10-30$ degrees, with the degree of misalignment being a strong function of mass. 
}

		Observationally, several studies have detected the intrinsic alignments for Luminous Red Galaxies (LRGs) to tens of Mpc scales \citep{Mandelbaum2006,Hirata2007,Okumura2009,Joachimi2011}, finding the radial 
		dependence of the signal to be consistent with the 
		linear alignment model. \cite{Hirata2007} and \cite{Joachimi2011} also observed luminosity dependence of \ia{}, with more luminous objects having stronger alignments. Many studies have also tried to measure \ia{} 
		within halos using large cluster samples \citep{Sifon2014,Schneider2013,Hung2012,Hao2011}, but so far, satellite alignments within groups and clusters have not been conclusively detected. \cite{Zhang2013} 
		used the Sloan Digital Sky Survey (SDSS) DR7 data to reconstruct the cosmic tidal field, and observed that the major axis of galaxies tends to be aligned with the direction of the filament or the plane of the sheet in which 
		they live. They also observed color, environment 
		and luminosity dependence of this alignment, with central, red and brighter galaxies showing stronger alignments.
		\cite{Blazek2012} and \cite{Chisari2014} also developed methods to  measure \ia{} using more general photometric samples from SDSS. 
		When using photometric redshifts, \ia{} measurements suffer from contamination from weak lensing. After accounting for possible weak lensing contamination, \cite{Blazek2012} and \cite{Chisari2014} found the \ia{} 
		signals for typical galaxies to be consistent with null detection.

		In this work, we will extend the study of \ia{} in LRGs to more faint objects using the low redshift (LOWZ)
		spectroscopic sample of LRGs from the SDSS-III BOSS survey. The LOWZ sample covers the same redshift range as the LRG sample used by \cite{Hirata2007}, but it extends to lower luminosities, 
		{with mean and median $r$ band magnitude fainter by $\sim0.3$ magnitudes in the redshift range that we use in this work. LOWZ also has higher comoving} 
		number density $\sim3\times10^{-4}h^{-3}\text{Mpc}^3$, an increase by a factor of 3.
		We test the validity of the non-linear alignment 
		model in both the linear and quasi-linear regimes using shape-density cross correlations for various samples defined using different galaxy properties.  We study the environmental 
		dependence of \ia{} by identifying galaxies in groups and study the \ia{} for satellites,
        BGGs (brightest group galaxies), and field galaxies {(galaxies in groups of
          multiplicity one in our sample)}. We also model the small-scale signal using a halo model fitting function 
		from \cite{Schneider2010}, and study the variations in small-scale intrinsic alignment amplitude with different galaxy properties.

		This paper is organized as follows: in section \ref{sec:formalism}, we describe the linear alignment model of \ia{} along with the halo model and the two point correlation functions we use to measure \ia{}. Section 
		\ref{sec:data} describes the SDSS-III BOSS LOWZ sample and selection criteria for the different sub-samples that we use. We present our results in section \ref{sec:results} and conclude in section 
		\ref{sec:conclusion}.
		
		Throughout we use standard cosmology with $h=0.7$, $\Omega_b=0.046$, $\Omega_{DM}=0.236$, $\Omega_\Lambda=0.718$, $n_s=0.9646$, $\sigma_8=0.817$ \citep[WMAP9,][]{Hinshaw2013}. All the distances are 
		in comoving \mpch, though $h=0.7$ as mentioned above was used to calculate absolute magnitudes and to generate predictions for the matter power spectrum. 
	
	\section{Formalism and Methodology}\label{sec:formalism}

		We model galaxy shapes as some purely random shape, plus two small but coherent shape distortions (``shears''), one due to intrinsic alignments (e.g., from some large-scale tidal field) and the other due
		to gravitational lensing.  The coherent part of the shears can be written as  $\gamma=\gamma^G+\gamma^I$, where $\gamma^G$ is the lensing shear and $\gamma^I$ is the 
		intrinsic shear. The two point correlation function for shear thus has the following contributions:
		\begin{equation}
			\langle\gamma\gamma\rangle=\langle\gamma^G\gamma^G+\gamma^{G}\gamma^I+\gamma^I\gamma^I\rangle=\xi_{GG}+\xi_{GI}+\xi_{II}
		\end{equation}
		$\xi_{GG}$ is the desired lensing signal, while the $\xi_{II}$ contribution comes from two nearby galaxies that are affected by the same tidal field and $\xi_{GI}$ contribution comes from the pair where 
		one galaxy shape is affected by the tidal field of the gravitational potential it lives in while a background galaxy is lensed by the same potential, see \cite{Hirata2004}. In Sec.~\ref{ssec:nla} we describe the linear 
		alignment 
		model used to model \ia{} at large scales, followed by the halo model prescription in Sec.~\ref{ssec:halo_model}, which describes the \ia{} at small scales.

		\subsection{Linear alignment model}\label{ssec:nla}
			One popular formalism to study intrinsic alignments at large scales is the linear alignment model \citep{Catelan2001,Hirata2004,Chisari2013}. In this section, we briefly describe the main 
			features of the model.
			
			The linear alignment model is based on the assumption that the intrinsic shear of galaxies is determined by the tidal field at the time of formation of the galaxy 
			\citep[assumed to be during matter domination;][]{Catelan2001}. Thus, we can write the intrinsic shear in terms of the primordial potential as 
			\begin{equation}
				\gamma^I=(\gamma^I_+,\gamma^I_\times)=-\frac{C_1}{4\pi G}(\partial^2_x-\partial^2_y,\partial_x\partial_y)\phi_p
				\label{eqn:gamma_phi}
			\end{equation}
			Using the form of $\gamma^I$ described in Eq.~\eqref{eqn:gamma_phi}, \cite{Hirata2004} derived the power spectrum for the two-point matter-\ia{} correlation functions, relating them to the linear matter power 
			spectrum, $P_\delta^{\text{lin}}$.
			\begin{align}
				P_{g+}(\vec{k},z)&=A_I b \frac{C_1\rho_{\text{crit}}\Omega_m}{D(z)} \frac{k_x^2-k_y^2}{k^2} P_\delta^\text{lin} (\vec{k},z)\label{eqn:LA+}\\
				P_{++}(\vec{k},z)&=\left(A_I \frac{C_1\rho_{\text{crit}}\Omega_m}{D(z)} \frac{k_x^2-k_y^2}{k^2} \right)^2P_\delta^\text{lin} (\vec{k},z)\label{eqn:LA++}\\
				P_{g\times}(\vec{k},z)&=A_I b\frac{C_1\rho_{\text{crit}}\Omega_m}{D(z)}\frac{k_x k_y}{k^2}P_\delta^\text{lin}(\vec{k},z)\label{eqn:LAx}
			\end{align} 
			Here, $P_{g+}$ ($P_{g\times}$) is the cross-power spectrum between the galaxy density field and the galaxy intrinsic shear along (at 45$^\circ$ from) the line connecting a pair of galaxies; $P_{++}$ is the shape-
			shape power spectrum for alignment along the line joining each pair of galaxies in the two-point correlations; and $b$ is the 
			linear galaxy bias. Following \cite{Joachimi2011}, we fix $C_1\rho_{\text{crit}}= 0.0134$, and use an arbitrary constant $A_I$ to measure the amplitude of \ia{} for different samples. $D(z)$ is the linear growth factor, 
			normalized to unity at $z=0$.
			
			To measure the two-point correlation functions, we cross-correlate two samples of galaxies, with one sample used as a biased tracer of the matter density (``density sample'', denoted $D$) and the other sample with 
			shapes to trace the intrinsic shear (``shape sample'', denoted $S$). See Sec.~\ref{ssec:corr} for more details. The galaxy bias entering Eq.~\eqref{eqn:LA+} is the bias of the density sample, while $A_I$ is 
			measured for shape sample.
			
			\cite{Bridle2007} suggested using the full non-linear matter power spectrum, $P_\delta^{\rm nl}$, in Eqs.~\eqref{eqn:LA+}--\eqref{eqn:LAx} to extend the linear alignment model to quasi-linear scales. This model is 
			called the non-linear linear alignment (NLA) model. While there are a number of reasons why this substitution of the non-linear matter power spectrum is unlikely to be valid to very small scales \citep[see discussion 
			in][]{Bridle2007}, it should allow at least some extension to smaller scales, and use of this model will permit us to compare our results more easily with many other works that have also adopted this model.
			In this work, we will use the non-linear matter power spectrum based on halofit model \citep{Smith2003}, generated with software CAMB \citep{Lewis2002}, with fixed WMAP9 cosmology from \cite{Hinshaw2013}.
			The expressions in Eqs.~\eqref{eqn:LA+}--\eqref{eqn:LAx} can be Fourier transformed to get the 3d correlation functions:
			\begin{align}
				\xi_{AB}(r_p,\Pi,z)=&\int \frac{\mathrm{d}^2k_\perp\mathrm{d}k_z}{(2\pi)^3}P_{AB}(\vec{k},z)\left(1+\beta_A\mu^2\right)\nonumber \\ &\left(1+\beta_B\mu^2\right)e^{i(r_p.k_\perp+\Pi k_z)}\label{eqn:xi}
			\end{align}
			Here $\mu=k_z/k$, and $\beta$ is the linear redshift distortion parameter\footnote{ We do not explicitly account for the Fingers of God effect due to the virial motions of galaxies within halos. However, these are 
			expected to be negligible in the projected correlations which are obtained 
			by integrating along the line-of-sight between $\pm100\mpch$.} 
			with the Kaiser factor $(1+\beta\mu^2)$
                        which accounts for the effects of
                        redshift-space distortions (RSD, see Appendix \ref{appendix:RSD}).
                        For quantities that include a galaxy
                          overdensity, the relevant $\beta$ is
                        defined as $\beta (z)=f(z)/b$, 
			where $f(z)$ is the logarithmic growth rate at redshift
			$z$ and $b$ is the galaxy bias of the sample. For the case of the shear, there is no RSD effect at leading order ($\beta_{+}=0$, see
                        Appendix \ref{appendix:RSD}). In $\Lambda$CDM, for a wide range of
                        redshifts, $f(z)\sim \Omega_m(z)^{0.55}$. {For the purpose of
                        modelling projected correlation functions, \citet{vdBosch:2013} show that the Kaiser approximation is sufficiently accurate.}
                        
			From data, we most easily measure $w(r_p)$, the 2d (projected) correlation function, which can be obtained by integrating the 3d correlation function over the line of sight separation $\Pi$
			\begin{align}
				w_{AB}(r_p)=&\int\mathrm{d}z \,W(z) \int\mathrm{d}\Pi\,\xi_{AB}(r_p,\Pi,z)\label{eqn:w}
			\end{align}
			Here $r_p$ is the 2d projected separation and $W(z)$ is the redshift window function {\citep{Mandelbaum2011}}:
			\begin{equation}
				W(z)=\frac{p_A(z)p_B(z)}{\chi^2 (z)\mathrm{d}\chi/\mathrm{d}z} \left[\int \frac{p_A(z)p_B(z)}{\chi^2 (z)\mathrm{d}\chi/\mathrm{d}z} \mathrm{d}z\right]^{-1}
			\end{equation}
			$p_A(z)$ and $p_B(z)$ are the redshift probability distributions for shape and density sample respectively and $\chi(z)$ is the comoving distance to redshift $z$.
			
			Doing the integral in Eq.~\eqref{eqn:w}, we get the 2d correlation functions:
			\begin{align}
				w_{g+}(r_p)=&\frac{A_I b_D C_1 \rho_{\text{crit}} \Omega_m}{\pi^2}\int\mathrm{d}z \frac{W(z)}{D(z)}\int_0^{\infty}\mathrm{d}k_z\int_0^{\infty} \mathrm{d}k_{\perp}\nonumber \\
				&\frac{k_\perp^3}{(k_\perp^2+k_z^2)k_z} P(\vec{k},z)\sin(k_z\Pi_\text{max})J_2(k_\perp r_p) \nonumber \\ 
                                &\left(1+\beta_{D}\mu^2\right) \label{eqn:wgp}
			\end{align}
                        where $b_D$ is the galaxy bias for density sample, while $\beta_{D}=f/b_D$. 
		 \wgx{} is expected to be zero by symmetry. The projected \ia{} auto-
			correlation function is defined as
			\begin{align}				
				w_{++}(r_p)=&\frac{\left(A_I C_1 \rho_{\text{crit}} \Omega_m\right)^2}{2\pi^2}\int\mathrm{d}z \frac{W(z)}{D(z)^2}\int_0^{\infty}\mathrm{d}k_z\int_0^{\infty} \mathrm{d}k_{\perp}\nonumber \\
				&\frac{k_\perp^5}{(k_\perp^2+k_z^2)^2k_z} P(\vec{k},z)\sin(k_z\Pi_\text{max})\times \nonumber \\ &[J_0(k_\perp r_p)+J_4(k_\perp r_p)] \label{eqn:wpp}
			\end{align}
			As discussed earlier, there is no RSD correction in
            $\gamma_I$ at leading order. Thus, the RSD correction\footnote{{This is different from equation C.1 in \cite{Blazek2011} and equation 12 in 
			\cite{Chisari2013}, which have an incorrect RSD factor in
            \wgp.} } in \wgp\ is only $(1+\beta_{D}\mu^2)$ (from the
        galaxy overdensity $\delta_D$) and there is no RSD correction in \wpp\ at leading order (see Appendix \ref{appendix:RSD} for 
			derivation).

			To measure the bias of the density sample and of the shape sample, $b_S$, we measure the two-point galaxy-galaxy cross correlation function
			{			
			\begin{align}
				&w_{gg}(r_p)=\frac{b_S b_D}{\pi^2}\int\mathrm{d}z \,{W(z)}\int_0^{\infty}\mathrm{d}k_z\int_0^{\infty}\mathrm{d}k_{\perp}\frac{k_\perp}{k_z} P(\vec{k},z)\nonumber \\ &\sin(k_z\Pi_\text{max})J_0(k_\perp 
				r_p)\left(1+\beta_S\mu^2\right)\left(1+\beta_D\mu^2\right)\label{eqn:wgg}  
			\end{align}
}

			\IA{} measurements that do not use spectroscopic redshifts suffer from gravitational lensing as a potential contaminant \citep{Joachimi2011, Blazek2012,Chisari2014}. Since we use 
			spectroscopic redshifts to select galaxy pairs, we expect negligible lensing contamination in our measurements from galaxy-galaxy lensing. Hence we do not include any contamination signals in our model.
			
		\subsection{Halo model}\label{ssec:halo_model}
			To extend this model of \ia{} into scales comparable to or smaller than the sizes of
            dark matter halos, \cite{Schneider2010} proposed the halo model for \ia{}. The halo
            model assumes that BCGs {(Brightest Cluster Galaxy)} or BGGs {(Brightest Group Galaxy)}
			are at the center of dark matter halos, with their shapes aligned with the host halos. Satellite galaxies tend to align radially with 
			the major axis pointing towards the center of the halo. In practice there can be substantial contamination in small scale \ia{} signal due to off-centering of BCGs, as well as misalignments between dark matter and 
			baryons. 

			\cite{Schneider2010} calculated the 1-halo \ia{} power spectrum using Monte Carlo simulations with the aforementioned complications, and they provide a fitting function for the halo model, given by
			\begin{equation}
				P_{\delta,\gamma_I}^{1h}=a_h \frac{(k/p_1)^2}{1+(k/p_2)^{p_3}}
				\label{eqn:halo_ps}
			\end{equation}
			The parameters $p_i$ are given by
			\begin{equation}
				p_i=q_{i1}\exp{\left(q_{i2}z^{q_{i3}}\right)}
			\end{equation}
			$z$ is the redshift and $q_{ij}$ are the parameters that we keep fixed to values described in Table~\ref{tab:haloparams}. Our choice of $q_{i2}$ and $q_{i3}$ values are the same as in \cite{Schneider2010}, while 
			the $q_{i1}$ are different, chosen to fit the shape of $w_{g+}$ for LOWZ sample. As in Sec.~\ref{ssec:nla}, we do the Fourier transform and line-of-sight integration to get
			\begin{equation}\label{eqn:halo_wgp}
				w_{g+}^{1h}=\int\mathrm{d}z \,W(z) \int\mathrm{d}\Pi \int \frac{\mathrm{d}^2k_\perp\mathrm{d}k_z}{(2\pi)^3}P_{\delta,\gamma_I}^{1h}(\vec{k},z)e^{i(r_p.k_\perp+\Pi k_z)}
			\end{equation}
			When computing the correlation functions from data, we use $\Pi \in [-100,100]\mpch$, which is much larger than a typical halo size. The $d\Pi$ integral can then be approximated as a delta function, and Eq.~
			\eqref{eqn:halo_wgp} can be written as 
			\begin{equation}\label{eqn:halo_wgp2}
				w_{g+}^{1h}=\int\mathrm{d}z W(z) \int \frac{\mathrm{d}^2k_\perp}{(2\pi)^2}P_{\delta,\gamma_I}^{1h}({k_\perp},z)e^{i(r_p.k_\perp)}
			\end{equation}
			\begin{table}
			%\centering
				\begin{tabular}{|c|c|c|c|}
				\hline
				Parameter Index & $q_{i1}$& $q_{i2}$& $q_{i3}$ \\ \hline
				1&0.005&5.909&0.3798\\
				2&0.6&1.087&0.6655 \\
				3&3.1&0.1912&0.4368 \\ \hline
				\end{tabular}
				\caption{Halo model parameters used in Eq.~\eqref{eqn:halo_ps}.}
				\label{tab:haloparams}
			\end{table}
			We emphasize that in order to fully explain and interpret the small-scale intrinsic alignments, a full halo model description would require us to describe the halo occupation statistics of each sample including 
			the Fingers of God effect, along with a 
			model for radial dependence of intrinsic shear and misalignment angles. Some of the assumptions made by \cite{Schneider2010} when computing the power spectrum might not be valid 
			in the context of some (or all) of our measurements. However, here we have a much simpler goal: we want to provide a simple fitting formula that approximately describes the observed intrinsic alignments of LOWZ 
			galaxies on small scales.  For this reason, we use the above formulae and fit for the amplitude 
			parameter $a_h$ as a simple way to encapsulate the 2-point correlation functions below $\sim 1.5\hmpc$ in terms of one number.  Thus, $a_h$ cannot be simply interpreted as a single quantity such as the 
			amplitude of a satellite galaxy radial shear.
		\subsection{Correlation function estimators}\label{ssec:corr}
			In order to compute the \ia{} signals, we cross-correlate the shapes of galaxies in the shape sample (S) with the positions of galaxies in the density sample (D). We use a generalized Landy-Szalay correlation 
			function estimator \citep{Landy1993} to calculate the cross-correlations 
%			\begin{align}
			\begin{equation}	
				\xi_{gg}=\frac{(S-R_S)(D-R_D)}{R_SR_D}=\frac{SD-R_SD-SR_D+R_SR_D}{R_SR_D},
			\end{equation}
			\begin{equation}	
				\xi_{g+}=\frac{S_+D-S_+R_D}{R_SR_D},
			\end{equation}
			\begin{equation}	
				\xi_{++}=\frac{S_+S_+}{R_SR_S}.
			\end{equation}
			$R_S$ and $R_D$ are sets of random points corresponding to the shape sample and the density sample, respectively. Here the terms involving shears for the galaxies are 
			\begin{equation}	
				S_+X=\sum_{i\in S, j\in X}\gamma_+^{(i)}\langle j|i\rangle
			\end{equation}
			and
			\begin{equation}	
				S_+S_+=\sum_{i\in S, j\in S}\gamma_+^{(i)}\gamma_+^{(j)}\langle j|i\rangle.
			\end{equation}
			 $\gamma_{+,\times}$ measure the components of the shear along the line joining the 
			pair of galaxies and at $45$ degrees from that line, respectively.
			 In our sign convention, positive $\gamma_+$ implies radial alignments, while negative $\gamma_+$ 
			implies tangential alignments. 

			Finally, the projection uses summation over bins in $\Pi$, 
			\begin{equation}	
				w_{ab}=\int^{\Pi_{\text{max}}}_{-\Pi_{\text{max}}}\xi_{ab}(r_p, \Pi)\mathrm{d}\Pi.
			\end{equation}
			We use $\Pi_\text{max}=100 \mpch$, with $d\Pi=10 \mpch$. Our choice is different from that used in \cite{Mandelbaum2006}, but our results and conclusions are not significantly affected by this difference, and when 				modeling the results we self-consistently include this choice in the model predictions.
			$w_{g\times}$, $w_{\times\times}$, and $w_{+\times}$ are defined in a similar fashion.  Note that our correlation 
			function estimator is different from those used by \cite{Joachimi2011}. Compared to the estimator 
			$DD/DR$, the Landy-Szalay estimator is less affected by geometrical effects from survey boundaries and holes, and it also minimizes the covariance matrix. 
			
			To get the error bars, we divide the sample {(described in Sec.~\ref{sec:data})} into 100 jackknife regions of approximately equal area on the sky, and compute the cross-correlation functions by excluding 1 
			jackknife region at a time.  { \cite{2009MNRAS.396...19N} have shown that for galaxy
            two-point correlation functions, which have a substantial contribution to the errors from
            cosmic variance, the jackknife covariance estimator tends to
            overestimate the covariances especially on small scales.  While the magnitude of the
            effect likely depends on the size of the jackknife regions compared to the scales used
            for measurements, it seems possible that our errorbars on the
            density sample bias will be somewhat 
            overestimated.  In contrast, \cite{Mandelbaum2005} showed that for two-point correlation functions
            that are dominated by shape noise, internal covariance estimates from the bootstrap
            agree with those from external methods at better than 20\%.  There is no reason to
            expect the jackknife (another internal covariance estimator) to behave differently in
            this particular case, so our error estimates on intrinsic alignments amplitudes should
            be quite reliable.}
						
			When the shape and density sample are the same, we fix the cosmology and then simultaneously fit $w_{gg}$ and $w_{g+}$ for the galaxy bias, $b$, and  \ia{} amplitude, $A_I$, for each jackknife sample. The final 				values for both parameters are the jackknife mean, with errors
			calculated from the jackknife variance. If the shape and density sample are different, the bias for the density sample ($b_D$) is measured separately from its auto-correlation function, and then is held fixed during the 				fitting procedure that uses the cross-correlation functions $w_{gg}$ and $w_{g+}$ to get the bias of the shape sample $b_S$ and its \ia{} amplitude $A_I$. This procedure leads to slightly 
			underestimated errors on our bias ($b_S$) measurements, but has a negligible effect on the \ia{} amplitude, since the errors on the \ia{} amplitude are dominated by shape noise. We also note that the size of our 
			jackknife regions on the sky is approximately $8$ degrees on a side, which corresponds to $\sim65\mpch$ at redshift of 0.16. We measure (and show) the correlation function out $\sim200\mpch$, but we fit the models 
			only for $r_p<65\mpch$ to avoid any issues resulting from the size of jackknife regions (most notably the error bars for $r_p>65\mpch$ will be underestimated). We test the effect of limiting our model to 
			$r_p<65\mpch$, by fitting out to $r_p\sim200\mpch$ and found that our results and conclusions remain unchanged. 

		\subsection{Weak lensing}\label{ssec:wl}
			To study the possible halo mass dependence of \ia{}, we estimate the halo mass of galaxies in our sample using galaxy-galaxy weak lensing, with our LOWZ galaxies as the lens sample. As described earlier, the 					observed 
			shear of a galaxy has two contributions, $\gamma=\gamma^G+\gamma^I$, where $\gamma^G$ is the tangential shear from lensing and $\gamma^I$ is the intrinsic shear of the galaxy. \cite{Blazek2012} used the 
			same SDSS source sample as used in this work to estimate the \ia{} contamination in galaxy-galaxy lensing (with SDSS LRG sample as lenses) and found that the \ia{} contamination is well below the statistical error 
			for this source sample. 
			Thus, for our lensing measurements we will assume that the contribution of the $\gamma^I$ term is well below the shape noise limit, and will ignore its contribution when modeling the lensing signal. 
			In the weak lensing regime, the observable is then tangential shear $\gamma_t=\gamma_t^G$,
			which relates to the surface density contrast $\Delta \Sigma (r_p)= \bar{\Sigma}(<r_p) - \Sigma(r_p)$ as 
			\begin{align}
				\gamma_t (r_p)=\frac{\Delta \Sigma (r_p)}{\Sigma_c}
			\end{align}
			$\Sigma_c$ is the critical surface density, given in comoving coordinates as
			\begin{equation}
				\Sigma_c=\frac{c^2}{4\pi G}\frac{D_s}{(1+z_l)^2D_l D_{ls}}
			\end{equation}
 			$z_l$ is the lens redshift; $D_s$ and $D_l$ are the angular diameter distances to the source galaxy and lens respectively while $D_{ls}$ is the angular diameter distance between the lens and source.
			
			To measure the lensing signal, we compute the surface density contrast around a lens galaxy using the statistic
			\begin{equation}
				\Delta \Sigma(r_p)=\frac{\sum_{ls}w_{ls}\gamma_t^{(ls)}\Sigma_c^{(ls)}}{\sum_{rs}w_{rs}}
			\label{eqn:wl}
			\end{equation}
			Here $w_{ls}$ is the weight for the lens-source pair,
			\begin{equation}
				w_{ls}=\frac{\Sigma_c^{-2}}{\sigma_\gamma^2+\sigma_{SN}^2}
			\end{equation}
			where the denominator includes the shape noise and measurement error for the galaxy shear estimates added in quadrature.  $w_{rs}$ is similar to $w_{ls}$, but with random points acting as the lens sample. Thus 
			dividing by $w_{rs}$ accounts for any galaxies that are physically-associated with the lens but are accidentally included in the source sample, which would lower the lensing signal if left uncorrected 
			\citep{Sheldon2004,Mandelbaum2005}. 
						
			Since we are primarily interested in measuring the average mass of galaxies within our
            sample, we only model the lensing using an NFW profile \citep{Navarro1996}, for
            $r_p<1\mpch$. The {generalized} NFW profile is given by {
			\begin{equation}
				\rho=\frac{\rho_s}{(r/r_s)^{-\alpha}(1+r/r_s)^{\alpha+\beta}}
			\end{equation}			}
			with projected surface density given by
			\begin{equation}
				\Sigma(r_p)=\int_0^{r_{\text{vir}}} \rho(r=\sqrt{r_p^2+\chi^2})d\chi.
			\end{equation}
			
			For the NFW profile, $\alpha=-1$ and $\beta=3$ are fixed. We define the concentration, $c_{180b}=r_{180b}/r_s$, and mass, $M_{180b}$, using a spherical overdensity of $180$ times the mean density:
            		\begin{equation}
				M_{180b}=\frac{4\pi}{3}r_{180b}^3 (180\bar{\rho}_m)
			\end{equation}
			
			Using the mass-concentration relation from \cite{2008JCAP...08..006M} in Eq.~\eqref{eqn:mass-concen}, we fit the signal for each jackknife sample in the range $0.05\mpch<r_p<0.3\mpch$. We limit the model to 
			$r_p<0.3\mpch$ to avoid contamination from the 1-halo satellite term and the halo-halo terms, which are important at larger $r_p$ values for galaxies in this mass range (see, e.g., figure 3 of 
			\citealt{2014MNRAS.437.2111V}, which shows the contributions of these terms for a lower mass sample; all contributions tend to shift to the right for higher mass). The final quoted 
			values for the mass are jackknife mean and standard error. When presenting mass results we will discuss sensitivity to this choice of mass-concentration relation:
			\begin{equation}
			c=5 \left(\frac{M_{180b}}{10^{14}h^{-1}M_{\odot}}\right)^{-0.1}
			\label{eqn:mass-concen}
			\end{equation}

	\section{Data}\label{sec:data}

		 The SDSS \citep{2000AJ....120.1579Y} imaged roughly $\pi$ steradians
		of the sky, and followed up approximately one million of the detected
		objects spectroscopically \citep{2001AJ....122.2267E,
		  2002AJ....123.2945R,2002AJ....124.1810S}. The imaging was carried
		out by drift-scanning the sky in photometric conditions
		\citep{2001AJ....122.2129H, 2004AN....325..583I}, in five bands
		($ugriz$) \citep{1996AJ....111.1748F, 2002AJ....123.2121S} using a
		specially-designed wide-field camera
		\citep{1998AJ....116.3040G} on the SDSS Telescope \citep{Gunn2006}. These imaging data were used to create
		the  catalogues of shear estimates that we use in this paper.  All of
		the data were processed by completely automated pipelines that detect
		and measure photometric properties of objects, and astrometrically
		calibrate the data \citep{2001ASPC..238..269L,
		  2003AJ....125.1559P,2006AN....327..821T}. The SDSS-I/II imaging
		surveys were completed with a seventh data release
		\citep{2009ApJS..182..543A}, though this work will rely as well on an
		improved data reduction pipeline that was part of the eighth data
		release, from SDSS-III \citep{2011ApJS..193...29A}; and an improved
		photometric calibration \citep[`ubercalibration',][]{2008ApJ...674.1217P}.

		\subsection{Redshifts}

		Based on the photometric catalog,  galaxies are selected for spectroscopic observation 
		\citep{Dawson:2013}, and the BOSS spectroscopic survey was performed
		\citep{Ahn:2012} using the BOSS spectrographs \citep{Smee:2013}. Targets
		are assigned to tiles of diameter $3^\circ$ using an adaptive tiling
		algorithm \citep{Blanton:2003}, and the data was processed by an
		automated spectral classification, redshift determination, and parameter
		measurement pipeline \citep{Bolton:2012}.

\begin{figure}
			\centering
			\includegraphics[width=\columnwidth]{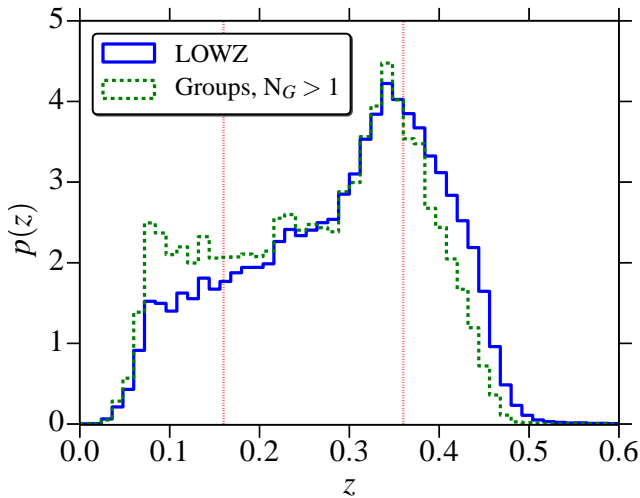} 
			\caption{Redshift distribution of LOWZ galaxies and our group sub-sample. Vertical lines mark the boundary of the redshift range that we use in this paper, $z=[0.16,0.36]$. 
			}
			\label{fig:z_hist}
		\end{figure}

		We use SDSS-III BOSS DR11 LOWZ galaxies, in the redshift range $0.16<z<0.36$. The LOWZ sample consists of Luminous Red Galaxies (LRGs) at $z<0.4$, selected from the SDSS DR8 imaging data and observed 
		spectroscopically in BOSS survey. Fig.~\ref{fig:z_hist} shows the normalized redshift distribution of the LOWZ galaxies. The sample is approximately volume limited in the redshift range $0.16<z<0.36$, with a number 
		density of $\bar{n}\sim 3\times10^{-4}h^3\text{Mpc}^{-3}$ \citep{Manera2014}. We 
		combine the spectroscopic redshifts from BOSS with galaxy shape measurements from \cite{Reyes2012}. BOSS DR11 has 225334 LOWZ galaxies within our redshift range. However, due to higher Galactic 
		extinction in some regions of the sky, \cite{Reyes2012} did not have shape measurements in those regions. After cutting out those regions, we are left with 173855 galaxies for our LOWZ density 
		sample, of which there are good shape measurements for 159621 galaxies, which constitutes our LOWZ shape sample. 
		\begin{figure}
			\centering
			\resizebox{1\columnwidth}{!}{\includegraphics[width=\columnwidth]{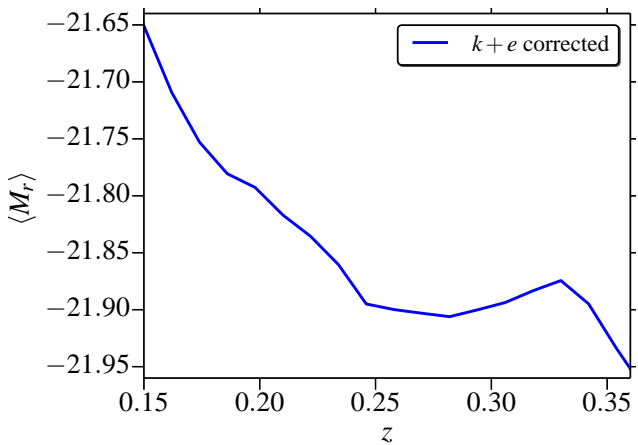} }
			\caption{Average $k+e$ corrected $r$-band absolute magnitude, $\langle M_r \rangle$, for the LOWZ sample as function of redshift. 
}
			\label{fig:mr_z}
		\end{figure}

		To compute absolute magnitudes, we use distances in the units of $\mpch$ for distance modulus computation and then apply $k+e$ corrections (to $z=0$) as described in \cite{Wake2006}. Fig.~\ref{fig:mr_z} 
		shows the average $k+e$ corrected magnitudes as function of redshift. Throughout, we will use the $k+e$ corrected magnitudes.

		To study the evolution of \ia{} with redshift and different galaxy properties, we cut the LOWZ sample into different sub-samples. Table~\ref{tab:samples} describes the selection criteria and various properties of different 
		sub-samples. Since our $k+e$ corrections are not perfect, we observe some redshift evolution in luminosity and color. To ensure that we select a fair samples based on luminosity and color cuts, those
		cuts are applied within redshift bins 
		where each bin has 10 per cent of the sample (equal sized bins in redshift percentile space). We also observe evolution of color with luminosity but we avoid selecting color samples from luminosity bins, as that changes 
		the redshift distribution ($\mathrm{d}n/\mathrm{d}z$) for different color samples, which complicates our correlation function calculations. For simplicity, we stick to redshift-based selection and keep the luminosity evolution in 
		mind when 
		interpreting the results for different color samples. Color and luminosity cuts are also based on percentile values. For example, the $L_1$ sample contains the galaxies in the $0-20^{th}$ percentile in luminosity within each 
		$z$ bin. $L_1-L_4$ samples go from brightest to faintest, with $L_1$--$L_3$ each containing 20\% of the galaxies, and $L_4$ having  the faintest 40 per cent of the galaxies ($L_4$ has more galaxies to get sufficient S/N). 
		 Similarly each color sample has 20 per cent of the galaxies and go from blue to red end, with $C_1$ being the bluest and $C_5$ being the reddest.

			\begin{table*}
				\begin{tabular}{|@{}c@{}|@{}c@{}|c|c|c|c@{}|c@{}|c|@{}c|c|c@{}|}%{|c @{} | c | @{}c @{}| @{}c @{}| @{}c @{}| c @{}| c @{}|}
					\hline
Dataset  & Cuts  & $N_D$  & $N_S$  & $\langle z \rangle$  & $\langle L_r/L_r^p \rangle$ & $\langle M_g$-$M_i\rangle$  & $\log\left(\frac{M_{180b}}{h^{-1}\Msun}\right)$  & $A_I$  & $b_S$  & $a_h$ \\ \hline
 Lowz &  & 173855 & 159621 & 0.28 & 0.95 & 1.18 & 13.18$\pm$0.05 & 4.6$\pm$0.5 & 1.77$\pm$0.04 & 0.08$\pm$0.01  \\ \hline 
BGG & $N_g>1$, BGG=True &19319 & 17916 & 0.27 & 1.31 & 1.16 & 13.5$\pm$0.1 & 8.1$\pm$1.5 & 2.54$\pm$0.13 & 0.41$\pm$0.05  \\ \hline 
Group & $N_g>1$& 44780 & 40928 & 0.27 & 1.01 & 1.19 & 13.3$\pm$0.1 & 3.6$\pm$1.0 & 2.67$\pm$0.14 & 0.26$\pm$0.03  \\ \hline 
Satellite & $N_g>1$, BGG=False & 25461 & 23012 & 0.27 & 0.78 & 1.22 & 13.1$\pm$0.1 & 0$\pm$1 & 2.77$\pm$0.15 & 0.16$\pm$0.04  \\ \hline 
Field & $N_g=1$& 129075 & 118693 & 0.28 & 0.93 & 1.18 & 13.13$\pm$0.06 & 5.0$\pm$0.6 & 1.46$\pm$0.07 & 0.011$\pm$0.007  \\ \hline 
Z1 & $0.16<z<0.26$ &67880 & 63180 & 0.21 & 0.91 & 1.17 & 13.17$\pm$0.06 & 4.1$\pm$0.8 & 1.66$\pm$0.07 & 0.05$\pm$0.01  \\ \hline 
Z2 & $0.26<z<0.36$ &105975 & 96441 & 0.32 & 0.98 & 1.19 & 13.18$\pm$0.07 & 5.1$\pm$0.8 & 1.88$\pm$0.05 & 0.16$\pm$0.02  \\ \hline 
$L_1$ &  $0\%<M_r<20\%$ &34760 & 31910 & 0.28 & 1.55 & 1.14 & 13.49$\pm$0.07 & 8.5$\pm$0.9 & 2.0$\pm$0.1 & 0.20$\pm$0.02  \\ \hline 
$L_2$ &  $20\%<M_r<40\%$ &34768 & 31910 & 0.28 & 1.04 & 1.16 & 13.25$\pm$0.09 & 5$\pm$1 & 1.74$\pm$0.08 & 0.09$\pm$0.02  \\ \hline 
$L_3$ &  $40\%<M_r<60\%$ &34768 & 31910 & 0.28 & 0.87 & 1.17 & 13.1$\pm$0.1 & 4.7$\pm$1.0 & 1.67$\pm$0.09 & 0.06$\pm$0.02  \\ \hline 
$L_4$ &  $60\%<M_r<100\%$ &69530 & 63830 & 0.28 & 0.65 & 1.23 & 12.96$\pm$0.09 & 2.2$\pm$0.9 & 1.70$\pm$0.08 & 0.03$\pm$0.02  \\ \hline 
$C_1$ &  $0\%<M_g-M_i<20\%$ &34760 & 31910 & 0.28 & 1.16 & 1.08 & 13.0$\pm$0.1 & 4.6$\pm$1.1 & 1.52$\pm$0.09 & 0.07$\pm$0.02  \\ \hline 
$C_2$ &  $20\%<M_g-M_i<40\%$ &34768 & 31910 & 0.28 & 0.99 & 1.16 & 13.24$\pm$0.10 & 5.0$\pm$1.0 & 1.72$\pm$0.09 & 0.06$\pm$0.02  \\ \hline 
$C_3$ &  $40\%<M_g-M_i<60\%$ &34768 & 31910 & 0.28 & 0.96 & 1.19 & 13.22$\pm$0.09 & 5.4$\pm$0.9 & 1.84$\pm$0.10 & 0.10$\pm$0.02  \\ \hline 
$C_4$ &  $60\%<M_g-M_i<80\%$ &34768 & 31910 & 0.28 & 0.92 & 1.21 & 13.36$\pm$0.09 & 5.8$\pm$1.0 & 1.95$\pm$0.10 & 0.10$\pm$0.02  \\ \hline 
$C_5$ &  $80\%<M_g-M_i<100\%$ &34760 & 31910 & 0.28 & 0.73 & 1.29 & 13.0$\pm$0.1 & 2$\pm$1 & 1.8$\pm$0.1 & 0.079$\pm$0.025  \\ \hline 						
				\end{tabular}

				\caption{Table describing various sub-samples in our analysis. `Cuts' describes the cuts implemented in our pipeline to select the sample, where $N_g$ is the group multiplicity. $N_D$ is the total number of 
				galaxies in the sample and $N_S$ is the number of galaxies in that sample with good shape measurements.  $\langle z \rangle$  and $\langle L_r/L_r^p \rangle$ are the average redshift and $r$-band absolute 
				luminosity for the sample, with $L_r^p$ being the pivot luminosity corresponding to r-band absolute magnitude $M_r^p=-22.0$. $M_{180b}$ is the average halo mass for the sample, measured from weak 
				lensing. \% denotes the percentile values for the sample. $b_S$ is the linear galaxy bias, and $A_I$ and $a_h$ are the \ia{} model amplitude for the 
				sample, calculated by cross-correlating the given shape sample with the full LOWZ sample as density sample, with its bias $b_D=1.77$ measured from LOWZ-LOWZ correlations and then held fixed. For joint 
				fitting \wgg{} and \wgp{}, we get reduced
                $\chi^2\in[0.8,2]$ for all samples presented here
                  with a probability to exceed this 
				$\chi^2$ by chance of $(0.77, 0.06)$ for the edges of this range.
				} 
				\label{tab:samples}
\end{table*}

			Due to the finite size of the optical fibers required to target galaxies for spectroscopy, the SDSS is not able to simultaneously take spectra for galaxy pairs separated by $\theta<62\arcsec$ on the sky. In regions of the 
			sky which are observed multiple times (due to overlapping tiles), a number of such cases can be resolved. Nevertheless a significant fraction of targets remain unassigned. This effect is known as fiber collisions. Since 
			fiber collisions occur preferentially in denser environments, the resulting incompleteness introduces a bias in correlation functions at all scales, with small scales being more strongly affected. 
			One way to correct for this, is to upweight the nearest neighbor of a fiber collided galaxy. This is based on the assumption that two very nearby red 
			sequence galaxies are likely to be within the same group or cluster.
			The nearest neighbor upweighting scheme has been shown to correct the bias in galaxy-galaxy correlation function on large scales ($\gtrsim$ fiber collision scale), but can introduce some bias at small scales 
			since the redshift separation between the collided pair of galaxies has been artificially set to zero, which is not always correct \citep{Reid2014}. Also while measuring the clustering of galaxies in subsamples, the 
			assumption that the fiber collided galaxy would also have been part of the same subsample as its nearest neighbor would be incorrect \citep[see e.g.,][in the context of SDSS-III CMASS galaxies]{More:2014}. 
			Therefore, instead of upweighting the neighbors, we include the fiber collided galaxies in our sample, but assign them redshifts of their nearest neighbors.
			Nevertheless, bearing in mind sample incompleteness and fiber collision 
			correction as possible source of systematics at small scales, we do not use the points below the fiber collision scale when fitting our models. 
			Throughout, the fiber collision scale will be marked with solid black lines at $r_p\sim 300$ kpc/h (fiber collision scale at $z=0.36$). In addition to the fiber collided galaxies, we also assign the nearest neighbor 
			redshifts 
			to galaxies for which the spectroscopic pipeline failed to obtain a redshift estimate. We calculate the absolute magnitudes along with $k+e$ corrections for these galaxies based on the nearest neighbor redshift 
			estimates.
		
		We use the sets of random points  provided by the BOSS collaboration. However, some of our subsamples have different redshift distributions from the full LOWZ sample, in which case we 
		generate new redshifts for the randoms from the subsample redshift distribution using the acceptance-rejection method. The acceptance-rejection method generates a set of random variates by uniformly sampling the 
		area of the probability distribution of the random variable. 
		
		\subsection{Shapes}\label{ssec:shapes}
			The catalogue of galaxies with measured shapes used in this
			paper (described in \citealt{Reyes2012} and further characterized in \citealt{Mandelbaum2013}) 
			was generated using 
			the re-Gaussianization method \citep{2003MNRAS.343..459H} of
			correcting for the effects of the point-spread function (PSF) on the
			observed galaxy shapes. The catalogue production procedure was described in detail in previous work, so
			we describe it only briefly here.  Galaxies were selected in a 9243
			deg$^2$ region, with an average number density of $1.2$ arcmin$^{-2}$.
			The selection was based on cuts on the imaging quality, data reduction
			quality, galactic extinction $A_r<0.2$ defined using the dust maps from
			\cite{1998ApJ...500..525S} and the extinction-to-reddening ratios from
			\cite{2002AJ....123..485S}, apparent magnitude (extinction-corrected
			$r<21.8$),  and galaxy size compared to the
			PSF.  The apparent magnitude cut used model
			magnitudes\footnote{\texttt{http://www.sdss3.org/dr8/algorithms/\\magnitudes.php\#mag\_model}}.   For comparing
			the galaxy size to that of the PSF, we use the resolution factor $R_2$
			which is defined using the trace of the moment matrix of the PSF
			$T_\mathrm{P}$ and of the observed (PSF-convolved) galaxy image
			$T_\mathrm{I}$ as
			\beq
			R_2 = 1 - \frac{T_\mathrm{P}}{T_\mathrm{I}}.
			\eeq
			We require $R_2>1/3$ in both $r$ and $i$ bands.
			
			The software pipeline used to create this catalogue obtains galaxy images in the $r$ 
			and $i$ filters from the SDSS `atlas images' 
			\citep{2002AJ....123..485S}.  The basic principle of shear measurement 
			using these images is to fit a Gaussian profile with elliptical
			isophotes 
			to the image, and define the components of the ellipticity
			\beq
			(e_+,e_\times) = \frac{1-(b/a)^2}{1+(b/a)^2}(\cos 2\phi, \sin 2\phi),
			\label{eq:e}
			\eeq
			where $b/a$ is the axis ratio and $\phi$ is the position angle of the 
			major axis.  The ellipticity is then an estimator for the shear,
			\beq
			(\gamma_+,\gamma_\times) = \frac{1}{2\cal R}
			\langle(e_+,e_\times)\rangle,
			\eeq
			where ${\cal R}\approx 0.87$ is called the `shear responsivity' and 
			represents the response of the distortion to a small 
			shear \citep{1995ApJ...449..460K, 2002AJ....123..583B}; ${\cal R} \approx 
			1-e_\mathrm{rms}^2$.  In the course of the re-Gaussianization
			PSF-correction method, corrections are applied to account for
			non-Gaussianity of both the PSF and the galaxy surface brightness
			profiles \citep{2003MNRAS.343..459H}.
			
			For this work, we do not use the entire source catalogue, only the
			portion overlapping the LOWZ sample.

			 When
			computing the intrinsic alignment correlation functions, we use the
			shear estimates from this catalog together with the redshifts from
			BOSS.  However, when computing the weak lensing signals around the
			LOWZ galaxies, we need estimates of the redshifts for the fainter
			source galaxies.   For this purpose, we
			use the maximum-likelihood estimates of photometric redshifts (photo-$z$) based on the five-band photometry from the Zurich Extragalactic
			Bayesian Redshift Analyzer \citep[ZEBRA,][]{2006MNRAS.372..565F}, which were
			characterized by \cite{2012MNRAS.420.3240N} and \cite{Reyes2012}.  In
			this work, we
			used a fair calibration sample of source galaxies with spectroscopic
			redshifts as defined in \cite{2012MNRAS.420.3240N} to calculate biases
			in weak lensing signals due to bias and scatter in the photo-$z$, and
			applied corrections that were of order 10 per cent ($\pm 2$ per cent)
			to the weak lensing signals.

		\subsection{Groups}\label{ssec:groups}

			We use the counts in cylinder (CiC) technique
                          \citep{Reid2009} to find the galaxies in groups. The CiC technique assumes that for galaxies in the same halos, the dominant 
			source of redshift separation between satellite and central galaxies are the line of sight relative velocities. Under such an assumption, a cylinder with size determined by the size and velocity dispersion of the halo 
			should be able to identify galaxy pairs within a single halo. \cite{Reid2009} calibrated the CiC technique for halos hosting LRGs using mock LRG catalogs and we use their dimensions for the cylinders, given by 
			$R_\perp <0.8 \mpch $  and $ |\Pi_\parallel | <20 \mpch $ in comoving coordinates. Once within-halo pairs are identified, we can use the friends-of-friends algorithm to find all the galaxies belonging to the same 
			group. 

\begin{figure}
				\centering
				\resizebox{1\columnwidth}{!}{\includegraphics[width=\columnwidth]{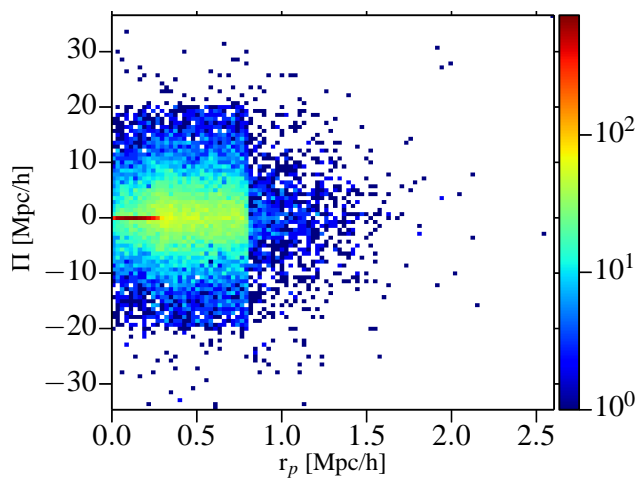} }
				\caption{The stacked distribution of satellite galaxies in our group sample, with respect to BGGs in the groups. The apparent sharp boundary at $r_p=0.8\mpch$ is due to the size of our CiC cylinders and the 
				fact that 
				our sample is dominated by two-galaxy (single satellite) groups. The strong peak at $\Pi=0$ and $r_p<0.3h^{-1}$Mpc arises primarily from the fiber collision corrected pairs.}
				\label{fig:group_dist}
			\end{figure}

			We note that the CiC parameters were optimized by \cite{Reid2009} for the SDSS LRG sample, and the LOWZ sample has a number density that is three times higher, extending to lower mass halos. Thus, using 
			the CiC parameters optimized for LRG sample can potentially lead to spurious group identification in our LOWZ group sample. To roughly quantify the level of contamination, we use the
			halo mass function of \cite{Tinker2008} and find that the lower limit of halo mass for a sample of halos with the LOWZ sample abundance is a factor of $\sim2.5$ less than a sample with the abundance of the LRG 				sample. Since the cylinder parameters should scale roughly like $M^{1/3}$, this will give $R_\perp 
			\lesssim0.6 \mpch $  and $ |\Pi_\parallel | \lesssim15 \mpch $ for the LOWZ sample. Using these parameters, we find that the satellite fraction in LOWZ reduces to 11.7 per cent, versus 14.6 per cent using the CiC 
			parameters from \cite{Reid2009}. Most of this difference is coming from the reduction in the $R_\perp$ size, since most of the satellites are within $|\Pi_\parallel | \lesssim10 \mpch$ as shown in 
			Fig.~\ref{fig:group_dist}.
			However, our calculation of CiC parameters for LOWZ is only approximate, as the CiC technique needs to be calibrated using mock catalogs, which is beyond the scope of this work. 
			Hence for our science analysis we will use the group catalog obtained using CiC parameters of \cite{Reid2009} { and we will discuss the impact of cylinder size on our conclusions in section \ref{ssec:systematics}}.

			Table \ref{tab:group_dist} shows the multiplicity function of our group sample and Fig.~\ref{fig:group_dist} shows the stacked 
			distribution of satellites with respect to BGGs, which we assume as the centre of groups. BGGs are selected as the brightest galaxy (in $r$ band) within the group.  The assumption that the brightest galaxy truly 
			resides at the center of the halo  fails for about 25\%-40\% of the groups within our
            mass range \citep{Skibba2011,2013MNRAS.435.2345H}, 
in which case the brightest galaxy is
            in fact a satellite  galaxy. {The mislabelling of the group center can suppress
              the radial alignment signal of satellite and also lead to spurious higher order
              moments in the measured correlation functions (see \citealt{Schneider2010} for an
              illustration of this effect)}. 
  {Also, in low multiplicity groups (which dominate our group sample) the misidentification
    of a central but non-BGG galaxy as satellite can enhance the satellite alignment signal (under
    the assumption that central galaxies are more strongly aligned with halo shape as traced by satellites).}
              These 
			potential sources of contamination should be kept in mind when interpreting our results on the environmental dependence of intrinsic alignments.
			\begin{table}
				\centering
			\hrule
				%\begin{multicols}{3}%{ \hfill \hfill }
				\begin{tabular}{@{}c@{}@{}c@{}@{}c@{}}\hspace{-20pt}%need hspace to undo mysterious right shift of the table
					\begin{tabular}{@{}c@{}@{}c@{}c@{}}
						$N_g$ & $N_G$ & $f$\\ \hline
1  &  130414  &  0.75  \\ 
2  &  14944  &  0.17  \\ 
3  &  2784  &  0.048  \\ 
4  &  780  &  0.018  \\ 
5  &  215  &  0.0062  \\ 
					\end{tabular}&
					\begin{tabular}{ |@{}c@{}|c@{}|c@{}| }
						$N_g$ & $N_G$ &$f$\\ \hline

6  &  84  &  0.0029  \\ 
7  &  33  &  0.0013  \\ 
8  &  17  &  0.00078  \\ 
9  &  7  &  0.00036  \\ 
10  &  2  &  0.00012  \\
					\end{tabular}&
					\begin{tabular}{ |@{}c@{}|c@{}|c@{}| }
						$N_g$ & $N_G$ &$f$\\ \hline
 
11  &  1  &  6.3$\times10^{-5}$  \\ 
12  &  1  &  6.9$\times10^{-5}$  \\ 
13  &  0  &  0  \\ 
14  &  1  &  8.1$\times10^{-5}$  \\ 
15  &  1  &  8.6$\times10^{-5 }$ \\ 
					\end{tabular}\hfill
				\end{tabular}
				%\end{multicols}
				\hrule	  
				\caption{Group multiplicity function for the group sample in redshift range, $0.16<z<0.36$. $N_g$ is the number of galaxies within a
                  group, $N_G$ is the total number of groups with $N_g$ galaxies, and $f$ is the fraction of the sample at a given group multiplicity $N_g$. 
                  }
				\label{tab:group_dist}
			\end{table}
	\section{Results}\label{sec:results}
		In this sections we present the results, beginning with correlation functions for full LOWZ sample and some tests for systematics. After that we primarily focus on evolution of \ia{} amplitudes with different galaxy 
		properties and end the section with some comparisons with other studies of \ia{}, using different \ia{} estimators where necessary. A summary of most of the results discussed in this section is presented in 
		table~\ref{tab:samples} and Fig.~\ref{fig:lowz_many}.
		\subsection{LOWZ sample}\label{ssec:results_lowz}
\begin{figure}
			\centering
			\resizebox{1\columnwidth}{!}{\includegraphics[width=\columnwidth]{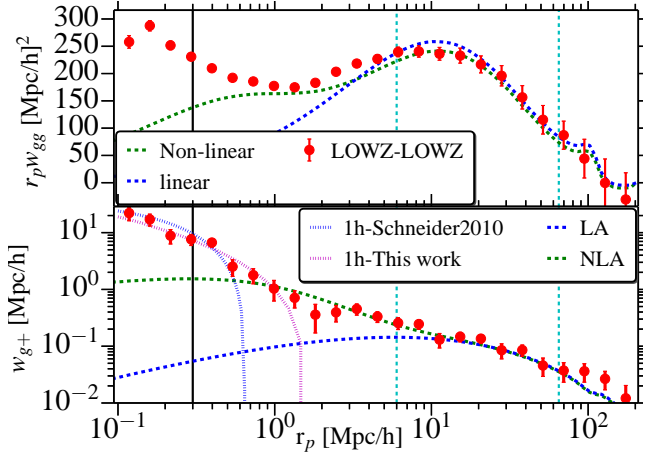}}
			\caption{The galaxy-galaxy correlation, \wgg{} (top, multiplied by $r_p$), and the density-shape correlation function, \wgp{} (bottom), for the full LOWZ sample in the redshift range $0.16<z<0.36$. The red points are the 
			measurements from the data, the 
			dashed green lines are the linear model with the non-linear matter power spectrum, and the dashed blue lines are the linear model with the linear matter power spectrum. The NLA model is fitted only in the range 
			$6\mpch<r_p<65\mpch$ (shown by a dashed vertical line), while the LA model is shown with the same parameters as the NLA model. The dotted red line in the bottom plot is the halo model fit to \wgp{} at small scales. 
			The black solid line shows the SDSS fiber collision scale at $z=0.36$.
			}
			\label{fig:lowz_w}
		\end{figure}

			We begin by showing our measurements for the full LOWZ sample. Fig.~\ref{fig:lowz_w} shows our results for the galaxy-galaxy correlation function \wgg{} and the density-shape correlation function, \wgp{}. 
			Throughout, when referring to 
			two point correlation measurements, our naming convention for different results will be in the form of ``Shape sample-Density sample". We fit these data to the following models: the galaxy-galaxy 
			correlation function calculated using non-linear matter power spectrum and linear galaxy bias, and the NLA model at $r_p>6$\mpch. For the full LOWZ sample, we find a bias $b=1.77\pm0.04$ and intrinsic 
			alignment amplitude $A_I=4.6\pm0.5$. The NLA model is a good fit to the \wgp{} measurement for $r_p>6\mpch$, and can be extended to $r_p\sim 4\mpch$, but there are significant deviations below that scale, 
			due to differing scale dependence.
 
			Our bias measurement appears to be in tension with results from \cite{Parejko2013}, who measured $b\sim2$ for the BOSS DR9 LOWZ sample. However, there are a number of differences between our 
			study and theirs: \cite{Parejko2013} (i) use a different redshift range, $z\in[0.2,0.4]$, (ii) use a smaller area DR9 sample, (iii) neglect redshift space distortions which can affect the large scale bias \citep[see e.g.,][]
			{vdB2013}, and  (iv) use a FKP weighting scheme which favorably weights the large bias high redshift galaxies. Using 
			the same redshift range as theirs, we also get a higher value of bias $b=1.85\pm0.04$. When RSD corrections in our model are neglected, we obtain an even higher value of $b=1.90\pm0.04$. We expect the 
			remaining discrepancy to be a result of the further differences in the weighting scheme and/or the sample size. 
			Our values are closer to the $b=1.85$ used by 
			\cite{Tojeiro2014} in the reconstruction scheme for the BAO measurement (\citealt{Tojeiro2014} use $\sigma_8=0.8$ and $n_s=0.95$).

		On non-linear scales, we use the halo model fitting formula for \ia{} defined in Sec.~\ref{ssec:halo_model}, fitting to the data with $0.3<r_p<1.5\mpch$. We find a good agreement with the data using the halo model 
		parameters as described in Table~\ref{tab:haloparams}, with halo model 
		amplitude $ a_h= 0.084\pm0.010$. Keeping $a_h$ fixed, but using fitting formula parameters from \cite{Schneider2010}, the model does not fit data well. This is not surprising as the halo model fitting function 
		does not explicitly model effects such as scale dependence of non-linear bias and \ia{} themselves, which can change both the amplitude and scale dependence of the signal. For this reason and those described in 
		Sec.~\ref{ssec:halo_model}, we caution against interpreting our $a_h$ values directly as an average \ia{} shear.
		
\begin{figure}
			\centering
			\resizebox{1\columnwidth}{!}{\includegraphics[width=\columnwidth]{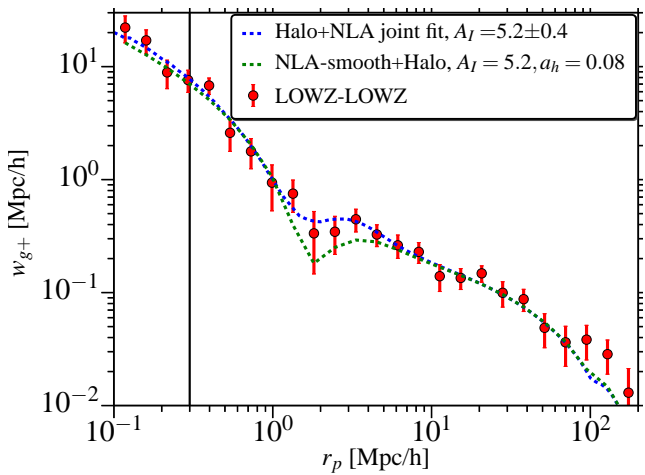}}
			\caption{{\wgp\ for the full LOWZ sample (same data as
                Fig.~\ref{fig:lowz_w}). Here we show the fits combining the halo model fitting
                function and the NLA model. The blue line is the joint fit ($0.3<r_p<65\mpch$) with
                an additional free parameter ($q_{21}$) in the halo model fitting function. The
                green line shows the combination (not a joint fit) of a smoothed NLA model with
                $k_{\text{smooth}}=1h/$Mpc and our usual halo model fitting function with best-fit parameters from Table~\ref{tab:haloparams} and Table~\ref{tab:samples}.}}
			\label{fig:lowz_wgp_joint}
		\end{figure}

		The halo model fitting formula combined with the LA model (linear alignment model with
        linear matter power spectrum) does not fit the data well in the intermediate regime
        $2<r_p<10\mpch$. {Fig.~\ref{fig:lowz_wgp_joint} shows the joint fit to the sum of
          the halo model fitting function and the NLA model in the range $0.3\mpch<r_p<65\mpch$. To
          better match the shape of the observed signal we let the parameter $q_{21}$ free in the
          halo model fitting function (in addition to the amplitude). We get best-fit values
          $A_I=5.2\pm0.4,a_h=0.014\pm0.004,q_{21}=1.1\pm0.1$. The higher value of $A_I$
          compared to the previous result, $4.6\pm 0.5$, is driven by the newly-included points
          below 6\mpch, while the value of $a_h$ is driven down by the small but non-negligible
          contributions from the NLA model below 1\mpch. The interpretation of this joint model is
          not clear, since it is possible that certain contributions to the signal are being double
          counted. Fitting our results to  the NLA plus halo model fitting function with free shape
          parameter will permit us to fit many of the signals across a range of scales, but it also
          complicates the interpretation of the amplitude variations that we study later (variations
          in $q_{ij}$ of the halo model fitting function also contribute to changes in $a_h$). For
          this reason, we do not show the combined fits for the NLA and halo model for other
          samples, but simply use the halo model fitting function as a convenient way to encapsulate
          the amplitude of the small-scale signal in a single number. For readers interested in
          self-consistent and uniform modeling using this ad hoc model at all scales, in
          Fig.~\ref{fig:lowz_wgp_joint} we also show the result of combining the one-parameter halo
          function with a smoothed NLA model \citep{Blazek2011} where the power spectrum in
          Eq.~\eqref{eqn:LA+} is modified with a smoothing function as shown in
          Eq.~\eqref{eqn:LA_smooth}, with smoothing scale $k_\text{smooth}=1h$/Mpc. Though not a very
          smooth function, this combination should provide a decent approximation to the measured
          signal at all scales as long  as the halo model fitting function is valid (i.e. density sample is LOWZ):
		\begin{equation}\label{eqn:LA_smooth}
			P_{g+}^{\text{smooth}}({k},z)=P_{g+}^{\text{NLA}}\exp{\left[-\left(\frac{k}{k_{\text{smooth}}}\right)^2\right]}
		\end{equation}
		}

\begin{figure}
				\centering
				\resizebox{1\columnwidth}{!}{\includegraphics[width=\columnwidth]{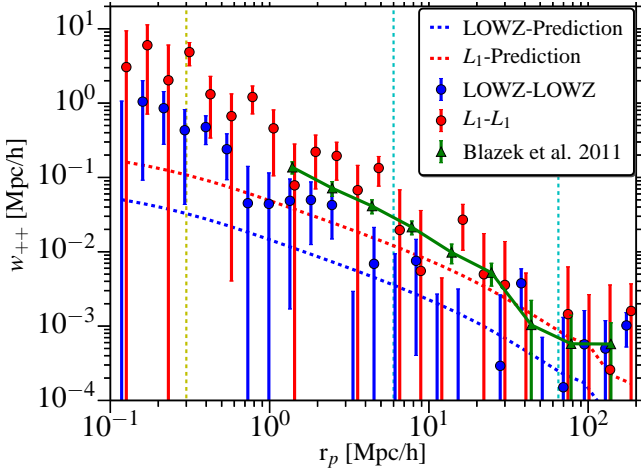}}
				\caption{ \wpp{} measurements for the full LOWZ sample (blue) and the $L_{1}$ sample (red). The green points show the \wpp{} measurement from \protect\cite{Blazek2011}. Our results are 
				dominated by shape 
				noise, and both the LOWZ and $L_1$ sample are consistent with a null detection at large scales. The dotted lines show NLA model predictions (using the best fit model to \wgp{}). Note that NLA fitting to 
				\wgp{} is only done for $6\mpch<r_p<65\mpch$.
				}
				\label{fig:lowz_wpp}
			\end{figure}

		Fig.~\ref{fig:lowz_wpp} shows our results for the $w_{++}$ measurement.  The NLA model with the best-fitting parameters from \wgp{} predicts a small but non-zero \wpp{}. Our signal is dominated by shape noise, and 
		our results (at $6\mpch<r_p<65\mpch$) marginally prefer the 
		null model (best fit $A_I=2\pm96$, $\chi^2=7.21$ with 8 bins, $\chi^2_0=7.22$,) over the NLA prediction using the amplitude from the fits to the \wgp{} signal ($\chi^2=8.0$) with a $\Delta \chi^2=0.8$, which is not a 
		statistically significant difference. Theoretically, from the NLA model with the best-fitting value of $C_1$ from the fits to \wgp, we expect $
		\wpp/\wgp\sim0.01$ (the ratio is scale dependent with a peak value of $\sim0.01$ for $r_p>6\mpch$). In the shape noise-dominated limit, in the case that the shape and density sample are the same, the standard deviations 
		of \wgp\ and \wpp\ are proportional to $\sigma_\gamma$ and $\sigma_\gamma^2$ \citep[see, e.g.,][]
		{Schneider2002}, respectively, where $\sigma_\gamma$ is the shape noise per component for our sample.  For this sample, $\sigma_\gamma\sim 0.2$.  
		Given a detection signal-to-noise ratio (SN) for \wgp\ on large scales of $9.2$ ($=A_I/\sigma_{A_I}$), the expected SN for \wpp\ on large scales is then $\text{SN}(\wpp)\sim 0.01 \,\text{SN}(\wgp)/0.2 \sim0.5$. Thus at 
		$r_p>6\mpch$ we expect to have a null detection of \wpp, consistent with our observations.  If we use scales below $6\mpch$, the $\chi^2=16.7$ for a null signal\footnote{All $p$-values in this paper calculated from either $
		\chi^2$ or $\Delta\chi^2$ use the simulation method described in \protect\cite{2004MNRAS.353..529H} to account for the fact that the jackknife covariances are noisy, which modifies the expected $\chi^2$ distribution.}, with 
		$9$ bins, giving a $p$-value$=0.1$ ($<2\sigma$).
		
		 Our \wpp{} results are inconsistent with those for SDSS-I/II LRGs from \cite{Blazek2011} ($\Delta \chi^2\sim11$, $p$-value$=0.004$), as shown in Fig.~\ref{fig:lowz_wpp}.
 		While calculating the projected correlation function from the 3d correlation functions in 
		 \cite{Okumura2009}, \cite{Blazek2011} assumed isotropic $\xi_{++}$ along with Gaussian 
		 and independent errors. These 
		 assumptions are likely to break down at some level, and can cause a potential discrepancy with our results.

 		Other possible sources of the discrepancy are as follows.  First, the LOWZ sample is fainter than the LRG 
		 sample in \cite{Blazek2011}. Since the 
		 \wpp{} signal goes as $A_I^2$, our expected signal is lower by a factor of $\sim4$ ($A_I^\text{LRG}=9.3$, $A_I^\text{LOWZ}=4.6$). 
		Our $L_1$ sub-sample is closer to the LRG sample in term of luminosity and the number of galaxies. However, as shown in Fig.~\ref{fig:lowz_wpp}, the large scale \wpp{} signal for the $L_1$ sample 
		is also consistent with a null signal (for $6\mpch<r_p<65\mpch$, the best fit $A_I=0\pm 44$ with $\chi^2=6.3$ for $8$ bins).  The $\Delta\chi^2$ when we compare this with the $\chi^2$ value for the value $A_I=9.22$ 
		from 
		\cite{Blazek2011} is only $\sim 0.5$.  Hence our $L_1$ sample results do not rule out the \cite{Blazek2011} results; however, what is interesting is the very large discrepancy in the significance of detection between the two 
		studies (with \citealt{Blazek2011} ruling out the null model at high significance). 

		A possible explanation for this discrepancy in detection significance comes from the use of different per-galaxy shear estimates.  The results in \cite{Okumura2009} that were used to present $w_{++}$ in \cite{Blazek2011} 
		used 
		isophotal shape measurements from the SDSS pipeline which are defined at a low surface brightness level and are not corrected for the effects of the PSF. PSF contamination and other systematics can  introduce 
		spurious shape correlations which can mimic a \wpp{} signal.
		However, it is still possible that we are seeing a real physical effect.  The SDSS isophotal shape measurements emphasize the outer parts of galaxies, while re-Gaussianization puts more weight on the inner regions of 
		galaxies. If the outer regions of galaxy shapes are more responsive to tidal fields, this could result in legitimately different \wpp{} signals when using the two different shear estimates. Using the MB-II SPH simulation,
		\cite{Tenneti2014b} found that differences in shape measurement methods, weighted towards outer or inner regions, leads to a difference of about 10\% in the \wgp\ measurement\footnote{The trend is mass-dependent, 
		and the difference is  larger for lower mass.  The number we quote here is for halo masses $M>10^{13}h^{-1}M_\odot$, which are the most appropriate ones for a comparison with massive galaxies like those in the LOWZ 
		sample.} {\citep[see also][]{Schneider2012,Allgood2006,Jing2002,Bett2012}}, which would translate to a $\sim20\%$ difference in 
		\wpp\ (with the right sign to explain what is seen here: \wgp\ is larger when weighting the outer regions more heavily). However, shape measurement methods in simulations may not correspond directly to shape 
		measurement methods in observations and hence there may not be a direct correspondence between results from 
		\cite{Tenneti2014b} and discrepancies observed between our results and that of \cite{Blazek2011}. We also note that \cite{Blazek2011} detected \wxx{} signal as well, but 
		our \wxx{} signal (not shown), like our \wpp{} signal, is consistent with zero. 
		A definitive answer to this puzzle requires deeper analysis of shape measurement methods and their impact on IA measurement, which we defer to future work. Due to the null detection of \wpp{} in the LOWZ 
		sample, we will not discuss \wpp{} hereafter.
			
		\subsection{Systematics tests}\label{ssec:systematics}
\begin{figure}
				\centering
				\resizebox{1\columnwidth}{!}{\includegraphics[width=\columnwidth]{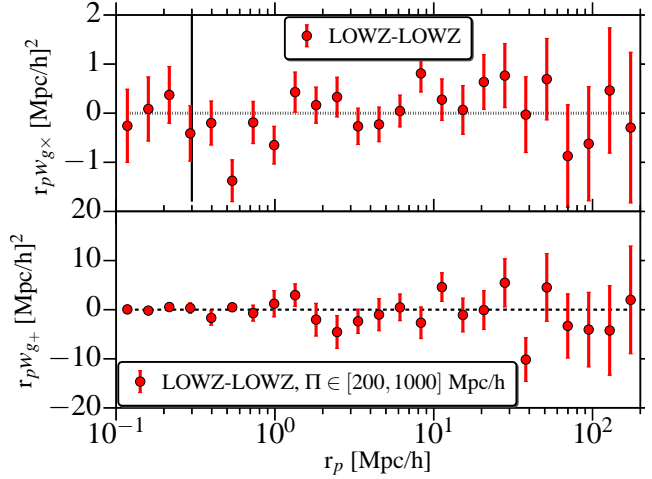}}
				\caption{\wgx{} and \wgp{} (at large $|\Pi|$) measurements for the full LOWZ sample. Both signal are consistent with zero. Note that both quantities are multiplied by $r_p$ in this plot. The increase in the statistical 
				errors on large scales is driven by large-scale shear systematics; see \protect\cite{Mandelbaum2013} for details.
				}
				\label{fig:lowz_sys}
			\end{figure}

			In this section, we present a \wgx{} measurement for the full LOWZ sample, and a \wgp{} measurement when integrating only over large $\Pi\in[200,1000]\mpch$ values. 
			Both these measurements are a good 
			test for systematics in our shape measurements 
			and correlation function calculations. From symmetry, \wgx{} is expected to be zero, while \wgp{} is also expected to have null signal when only large $\Pi$ values are considered since the galaxy pairs are too far 
			apart along the line-of-sight. As shown in Fig.~\ref{fig:lowz_sys}, the \wgx{} signal is consistent 
			with zero, with $\chi^2_\text{red}=1.48$ for $25$ bins (across all scales shown on the plot) and $p$-value$=0.34$. \wgp{} at large 
			$|\Pi|$ values is also consistent with zero, with $\chi^2_\text{red}=1.41$ and $p$-value$=0.40$.  We therefore conclude that these signals, which could reveal possible systematic errors, are fully consistent with zero.  
			There are also no patterns evident in the residuals from zero, just a slight hint that errorbars might be underestimated.
			
\begin{figure}
				\centering
				\resizebox{1\columnwidth}{!}{\includegraphics[width=\columnwidth]{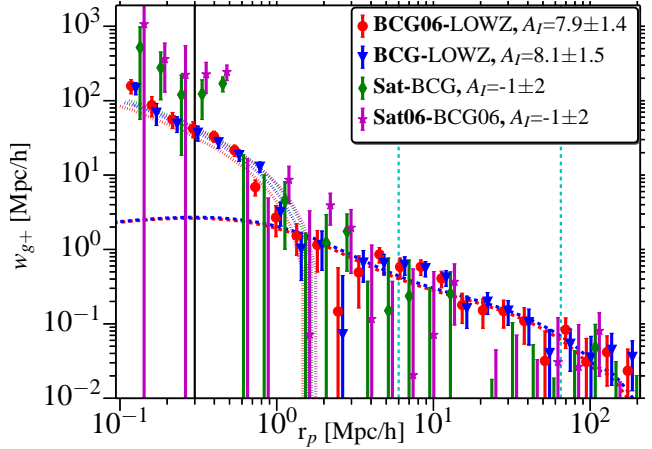}}
				\caption{{Comparison of \wgp\ from two group sample with different CiC
                    cylinder sizes. BCG-LOWZ (Blue) and Sat-BCG (green) correlations are for the group sample defined in Sec.~\ref{ssec:groups}, while BCG06-LOWZ and Sat06-BCG06 are for group sample with cylinder size $r_p<0.6\mpch$ and $|\Pi|<15\mpch$. We find consistent results for both group samples. }}
				\label{fig:bcg06}
			\end{figure}

			{Fig.~\ref{fig:bcg06} shows the comparison of \wgp\ for two group samples with
              different CiC cylinder sizes. To test for the effect of our cylinder size choice, we
              make another group sample with cylinder size $r_p<0.6\mpch\ $ and $|\Pi|<15\mpch\ $
              (see Sec.~\ref{ssec:groups} for reasons to choose alternative cylinder sizes) and
              compare the results with our main group sample as defined in
              Sec.~\ref{ssec:groups}. We find consistent values (well within $1\sigma$) for the
              linear bias and \ia\ amplitudes $A_I$ and $a_h$ for all the samples used here (group,
              BGG, satellite, field). We do observe some differences in the small-scale clustering
              signal, which is probably due to some contamination of our main group sample by
              non-associated pairs. This does not affect any of our science results, and since the alternate group sample has not been validated with simulations, we will use the main group sample as defined in Sec.~\ref{ssec:groups} for our science results.
			}


		\subsection{Luminosity Dependence}\label{ssec:result_lum}		

\begin{figure}
				\centering
				{\includegraphics[width=\columnwidth]{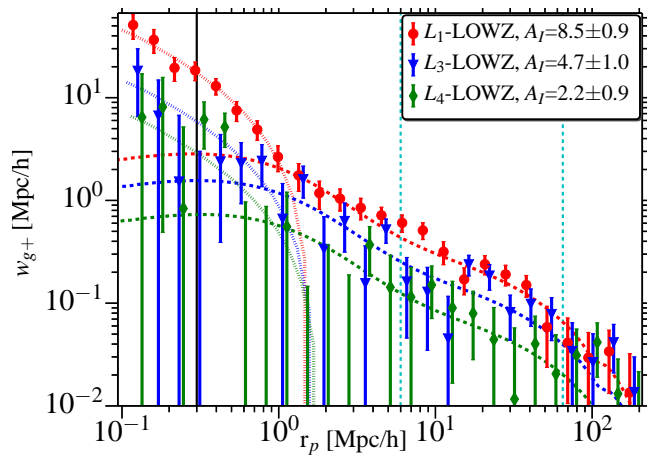} }
				\caption{Density-shape correlation functions for samples defined using luminosity cuts, $L_1$, $L_3$, $L_4$ samples, as specified in Table~\ref{tab:samples}. $L_1-L_4$ are arranged in order of luminosity, with 
				$L_1$ being brightest and $L_4$ being faintest. The $L_2$ sample is not shown in the figure for clarity. Brighter galaxies have higher \ia{} amplitude, with luminosity trend of $A_I$ being well described by a power 
				law (Eq.~\ref{eqn:lum_powerlaw}).
				}
				\label{fig:lum_w}
			\end{figure}

			Previous studies using LRGs have observed luminosity dependence of \ia{} \citep{Hirata2007,Joachimi2011}, where brighter galaxies tend to have higher \ia{} amplitude. LOWZ sample allows us to study luminosity 
			dependence by going to fainter luminosities, with spectroscopic redshifts for all galaxies. 
			To study luminosity dependence, we divide the LOWZ sample into four sub-samples, based on $r$-band absolute magnitude $M_r$. The sample are defined according $M_r$ percentiles within $z$ bins (each $z$ 
			bin has 10\% of the sample). $L_1$ has the brightest 20\% of the galaxies, followed by $L_2$ and $L_3$ samples which have next 20\% galaxies each based on $M_r$. $L_4$ has the faints 40\% of the galaxies. 
			Due to the decrease in signal to noise at fainter end, we do not cut $L_4$ sample into more sub-samples.
			Our density-shape correlation function measurements for the luminosity samples are shown in Fig.~\ref{fig:lum_w}. All four samples are cross correlated with LOWZ as the density sample, and the LOWZ bias is 
			fixed to $b_D=1.77$ (Sec.~\ref{ssec:results_lowz}) when modeling the results.
			Fits to $w_{gg}$ (not shown) reveal that the $L_1$ sample has the highest bias, followed by the $L_2$, $L_3$ and $L_4$ samples though differences in last three samples are statistically not very significant(see 
			Table~\ref{tab:samples} for values of fit 
			parameters like the bias). However, the overall trend in the evolution of bias with luminosity is consistent with the fact that brighter galaxies live in more massive and hence more biased halos. 
			
			We also see luminosity variation of the \ia{} amplitude and therefore of the NLA model best-fitting amplitude $A_I$, with brighter galaxies having higher amplitude. Since the density sample is the same in each case, 
			the increasing $w_{g+}$ amplitude reflects the different \ia{} for the different luminosity samples.  Following \cite{Joachimi2011}, we use a power law function to study the variation of $A_I$ with luminosity.
			\begin{equation}
				A_I(L_r)=\alpha\left(\frac{L_r}{L_r^p}\right)^\beta \label{eqn:lum_powerlaw}
			\end{equation}
			$L_r$ and $M_r$ are the $r$-band luminosity and absolute magnitude, respectively, and $L_r^p (M_r^p)$ is a pivot luminosity (magnitude), chosen to be $M_r^p=-22.0$. Fitting to the samples defined by our 
			luminosity cuts, we 
			find $\alpha=4.9\pm0.6$ and slope $\beta=1.30\pm0.27$ (where these errors come from the jackknife). Our results are quite consistent with those of \cite{Joachimi2011}, who found $\alpha=5.76^{+0.60}_{-0.62}$ and $
			\beta=1.13^{+0.25}_{-0.20}$ (using MegaZ-LRG + SDSS LRG + L4 + L3), with the same k-corrections and choice of $M_r^p$.  
			
			On small scales, we also see luminosity evolution of the halo model fitting function amplitude $a_h$, with brighter galaxies having higher amplitude. Fitting a power law similar to Eq.~\eqref{eqn:lum_powerlaw}, we 
			find a power-law amplitude  $\alpha_h=0.081\pm0.012$, and a power-law index $\beta_h=2.1\pm0.4$ (subscript ``h'' denotes fits to halo model fitting function parameters). The scaling of $a_h$ with 
			luminosity is somewhat steeper than that of $A_I$. This difference is likely due to that fact that more luminous galaxies live in denser regions, where tidal fields are stronger and hence leads to stronger \ia{} {\citep{Pereira2008}}. This will 
			increase \wgp{} at all scales, but the impact will be much stronger at smaller scales, leading to stronger scaling of $a_h$. This effect also contributes to the change in shape of  
			\wgp{} for $0.3<r_p<10$\mpch{} as the luminosity varies.
			\begin{figure*}
				\includegraphics[width=1.0\textwidth]{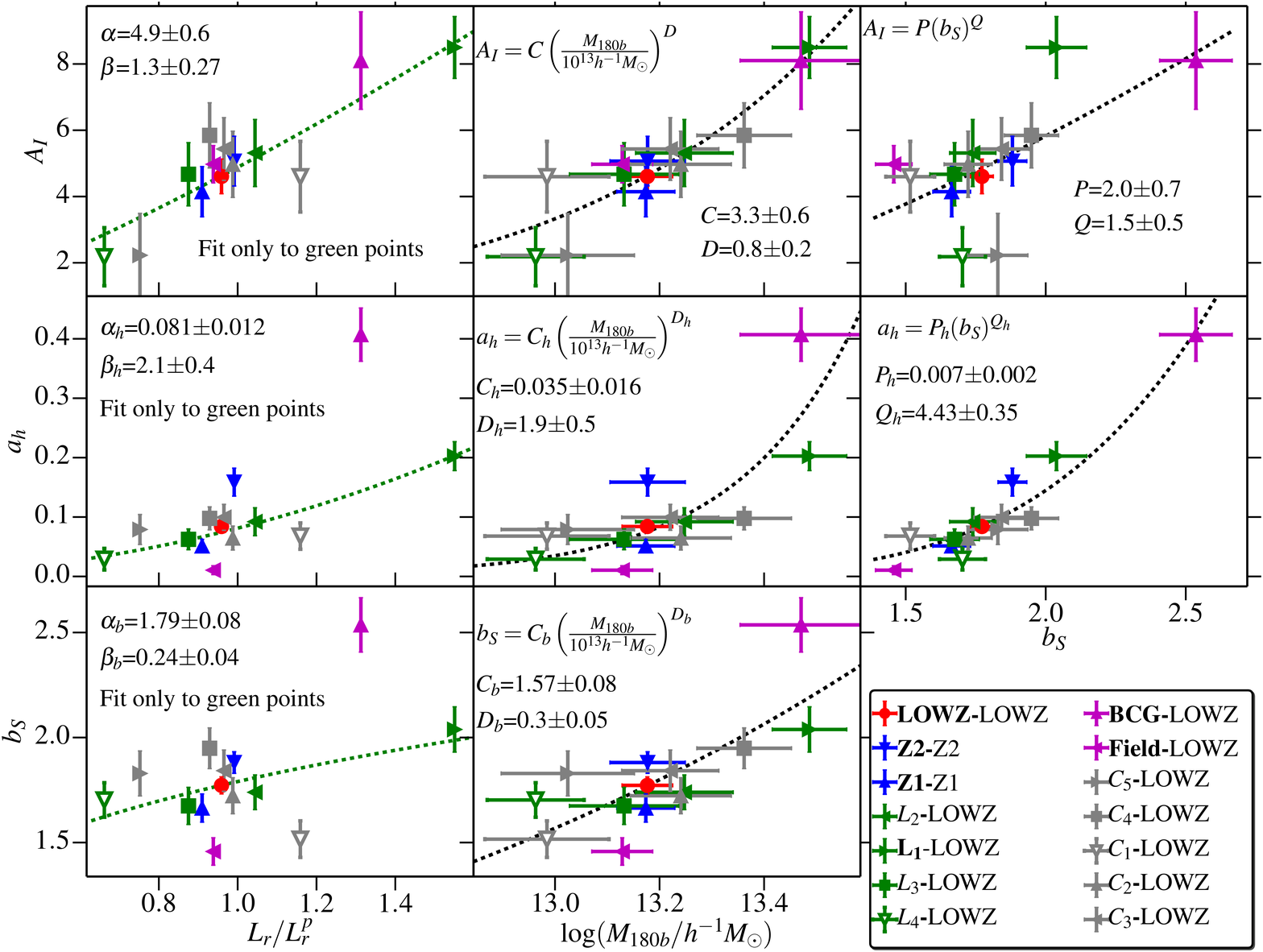}
				\caption{
				\IA{} amplitudes and bias for various shape samples, as a function of different galaxy properties of the shape sample. Note that the density sample is fixed to full LOWZ sample for all points, except for Z1 
				and Z2, for which density sample are Z1 and Z2 themselves, respectively. The top row shows $A_I$ as a function of different properties of shape sample. $A_I$ shows clear evolution with luminosity 
				($M_r$) as well as galaxy 
				mass and bias, with negligible evolution in redshift. The black dotted line in $A_I$ vs. r$M_r$ shows a power law fit to the luminosity samples (green points). Similarly in $A_I$ vs $\log(M_{180b}/h^{-1}\Msun)$ 
				($M_{180b}$ is the halo mass from weak lensing), the black dotted line is the power law fit,
				 using all the points. The middle row shows the halo model amplitude, $a_h$, as a function of different galaxy properties. For cases where the density sample is fixed to LOWZ, the effects of the non-linear bias of 
				 the density sample is the same. Therefore, the observed dependence of $a_h$ is likely due to the dependence of \ia{} on the properties of the shape sample. We find tight correlation between large scale bias of 
				 shape sample and $a_h$, which could partially arise from the dependence of the halo model amplitude on the non-linear bias of the shape sample (see Sec.~\ref{ssec:result_env} for more detailed discussion).
				 The black dotted line in $a_h$ vs $M_r$ is the power law fit to luminosity samples (green points), and the dotted line in $a_h$ vs $b_S$ is the power law  fit using all the points. Please see 
				 Sec.~\ref{ssec:wl_results} for possible systematics in the $a_h$ vs. mass relationship.  We emphasize that galaxy properties 
				 shown on the $x$-axis are correlated, for example, more luminous galaxies also have higher bias and live in more massive halos. Due to this correlation, we do not attempt to model \ia{} amplitudes by 
 				 simultaneously using more than one such property. 
				}
				\label{fig:lowz_many}
			\end{figure*}
		\subsection{Redshift Dependence}\label{ssec:results_z}
			To study the redshift dependence of \ia{}, we divide our sample into two redshift bins, Z1: $0.16<z<0.26$, Z2: $0.26<z<0.36$. 
			Due to the small redshift range of the LOWZ sample, we cannot cut our samples further, as 
			the edge effects from the redshift boundaries can then introduce significant bias in the correlation functions.
			
\begin{figure}
				\centering
				\resizebox{1\columnwidth}{!}{\includegraphics[width=\columnwidth]{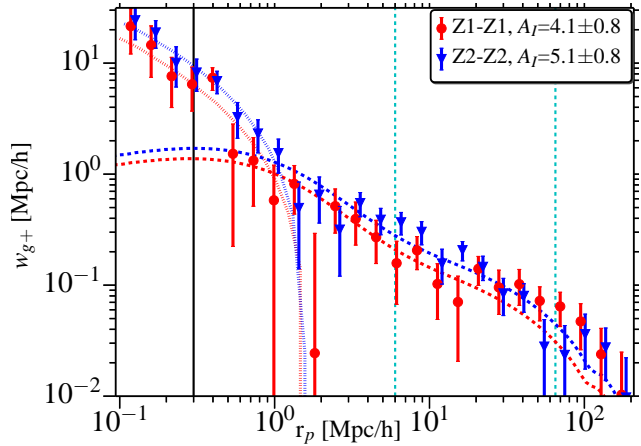} }
				\caption{Density-shape correlation functions for samples defined with redshift cuts, Z1 ($0.16<z<0.26$) and Z2 ($0.26<z<0.36$). We do not find any statistically significant redshift evolution for \ia{}.}
				\label{fig:z_w}
			\end{figure}

			Fig. \ref{fig:z_w} shows the density-shape correlation measurements for both of the redshift samples. 
			From \wgg{} measurement (not shown), we find higher galaxy bias for Z2 sample ($b=1.88\pm0.05$) compared to Z1 sample ($b=1.66\pm0.07$), which is consistent with the fact that among halos of similar mass, 
			ones at higher redshift are more 
			biased. We do not find any redshift evolution for $A_I$, which is consistent with the {LA} paradigm, where galaxy alignments are set at the time of galaxy formation and hence we do not expect any significant 
			redshift evolution for \ia{} beyond what is implicitly included due to use of a redshift-dependent nonlinear matter power spectrum that determines the gravitational tidal field.  However, given the short redshift 
			baseline, our power to constrain evolution of \ia{} with redshift is limited.
			
			On the small scales where we fit the halo model fitting function, there are significant differences between the two redshift ranges. The Z2 sample has a higher halo model amplitude. However, the Z2 sample is also 
			expected to have a higher non-linear bias which will 
			contribute to the increase in $a_h$. Processes within groups and clusters, such as galactic mergers, stripping, peculiar motion and tidal torquing will also play an important role in determining 
			the small scale signal {\citep[see, e.g.][]{Pereira2008}} and can contribute to the redshift evolution of $a_h$. Due to a degeneracy between \ia{} amplitude and non-linear bias within the halo model, we cannot separate out the effects of these 
			processes.

		\subsection{Color Dependence}\label{ssec:results_color}
\begin{figure}
				\centering
				\resizebox{1\columnwidth}{!}{\includegraphics[width=\columnwidth]{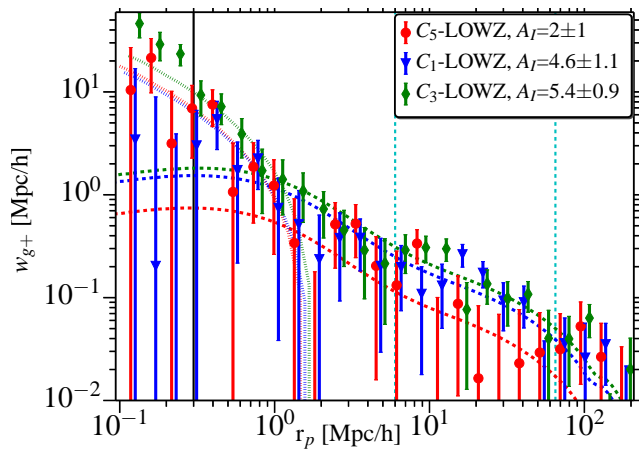} }
				\caption{Density-shape correlation functions for samples defined with color cuts, with $C_1-C_5$ arranged from bluest to reddest sample. We observe some \ia{} variations between different samples, but these 
				variations are explained by the luminosity difference between the different samples (see Fig.~\ref{fig:lowz_many}).}
				\label{fig:color_w}
			\end{figure}

		Our sample consists of only luminous red galaxies, so we cannot divide it into subsamples to study difference in \ia{} for red and blue galaxies. However, we do split our sample based on $M_g-M_i$ color to study 
		\ia{} dependence on color within the LRG sample. We divide our sample into five sub-samples based on color, with each sample having 20 per cent of the LOWZ galaxies. Color cuts were applied in $z$ bins to take out the 
		$z$ evolution 
		of color and make sure we select a fair sample. The five samples, $C_1-C_5$ are arranged from the bluest to reddest. We observe significant luminosity evolution across the five samples, with redder samples getting 
		progressively fainter. We do observe some \ia{} evolution across the different samples, but find that the evolution can be well explained by the luminosity variations alone (see Fig.~\ref{fig:color_w}). We note that the $C_1$ 
		sample 
		is expected to have some contamination from late-type galaxies \citep{Masters2011}, 
		but we do not observe a very significant deviation in \ia{} signal from the expectation based on luminosity or mass scalings. 
		
		Our results suggests that luminosity and mass are more important properties in determining \ia{} signal for red sequence galaxies, with color variations being absent or at least sub-dominant. This is an important test of the 
		models that are commonly used to predict the \ia{} contamination of future weak lensing surveys \citep[e.g.,][]{Joachimi2011}, which split galaxies into a red and a blue sample without 
		permitting any variation in \ia{} with color within those two samples.  Our results validate this choice for the red galaxies, which are currently more important since they are the ones for which there is a robust \ia{} 
		detection.

		\subsection{Weak Lensing}\label{ssec:wl_results}
\begin{figure}
				\centering
				\resizebox{1\columnwidth}{!}{\includegraphics[width=\columnwidth]{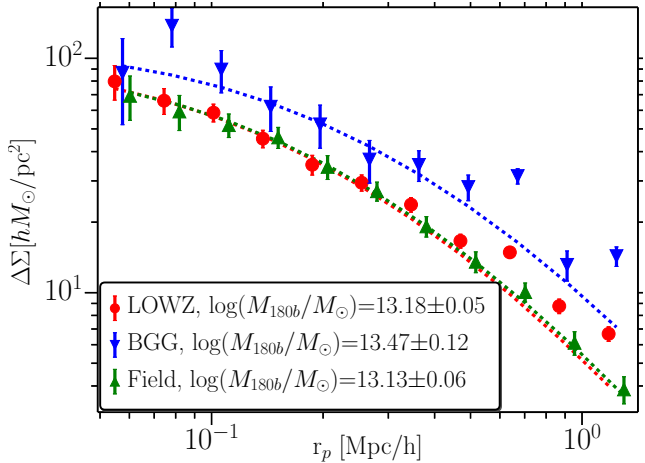}}
				\caption{Weak lensing measurement for three different samples: LOWZ, BGGs, and field galaxies as indicated in the legend. The dashed lines are the NFW profile fits to the signals. Deviations from the NFW in 
				case of BGGs and LOWZ are due to the presence of satellites in these samples (even the BGG sample, due to the fact that selecting a BGG purely based on luminosity is imperfect).  The field galaxy sample is  
				the least likely to suffer from satellite contamination, so its signal is well described by the NFW profile. 
				}
				\label{fig:wl_signal}
			\end{figure}

			To study the halo mass dependence of \ia{}, we compute the average halo mass of galaxies within different samples using galaxy-galaxy lensing. 
			Figure~\ref{fig:wl_signal} shows the weak lensing signal for LOWZ, BGG and field galaxies, with points being 
			measurements from data and dashed lines are the NFW profile fits, with concentration fixed using concentration-mass relation defined in Eq.~\eqref{eqn:mass-concen}. The signal for BGGs deviates from an 
			NFW profile for $r_p\gtrsim0.5\mpch$. These deviations could be due to some satellite contamination in our BGG sample, which is expected at the level of tens of per cent. At small scales, BGG mis-centering effects 
			(which we have not accounted for in our fits) can also lead to deviations from the NFW profile resulting in underestimated halo mass. 
			
			To check for the effect of using fixed concentration-mass relation defined in Eq.~\eqref{eqn:mass-concen}, we vary the amplitude in Eq.~\eqref{eqn:mass-concen} by $20\%$ and re-fit the signal from field sample, using 
			$0.05\mpch<r_p<1.0\mpch$ (different from range used for main results). We find that the final mass measured changes by $\sim -12\%$.

			Mass measurements for different sample are given in table.~\ref{tab:samples}. For full LOWZ sample, we get $\log(M/h^{-1}\Msun)=13.18\pm0.05$ and for our brightest sample, $L_1$, we get $\log(M/
			h^{-1}\Msun)=13.49\pm0.07$. Using the clustering analysis, \cite{Parejko2013} found $\log(M/h^{-1}\Msun)\sim13.72$ for the DR9 LOWZ sample. Their masses correspond to FOF halos with a linking length of 
			0.2 which are expected to have a higher overdensity \citep[see][]{More2011}, worsening the discrepancy we see here. However, clustering mass estimates are obtained in a less direct way that can 
			have some bias due to modeling assumptions. Clustering and galaxy-galaxy lensing also have different redshift window functions and \cite{Parejko2013} used a different redshift range, $z\in[0.2,0.4]$, due to which it 
			is hard to do a fair comparison between the two studies.

			The {second} column in Fig.~\ref{fig:lowz_many} shows the variations in \ia{} amplitudes and galaxy bias as function of mass. As expected, more massive galaxies have higher bias, with the mass dependence of the 
			bias being well-described by a power law, $b\propto M_{180b}^{0.30\pm0.05}$. \IA{} amplitudes are also strongly 
			correlated with the mass, with more massive galaxies showing stronger \ia{}. The relation between $A_I$ and mass can be well defined by a power law:
			\begin{equation}
				A_I=C\left(\frac{M_{180b}}{10^{13}h^{-1}\Msun}\right)^{D}
			\end{equation}

			We find an amplitude $C=3.3\pm0.6$ and index $D=0.8\pm0.2$. 
			Similarly for the small scale amplitude $a_h$, we get $C_h=0.035\pm0.016$ and index $D_h=1.9\pm0.5$. However, this fit is significantly effected by the BGG-LOWZ correlation point, where BGG mass measurement 				are affected by satellite contamination and BGG mis-centering effects. Excluding the BGG-LOWZ point, we find $C_h=0.047\pm0.012$ and index $D_h=1.2\pm0.3$. 
			We also note that we can only test a small range in mass using our LOWZ sample, and this result is unlikely to be valid when extrapolated significantly outside this mass range.
			
		\subsection{Environment Dependence}\label{ssec:result_env}
			One of the main goals of this work is to address the environmental dependence of intrinsic alignments. The halo model of intrinsic alignments \citep{Schneider2010} predicts that satellite galaxies tend to align 
			radially within their host halos, while the BGGs tend to align with the halo shape, which itself tends to align with the large scale tidal gravitational field according to the linear alignment model. As described in Sec.~
			\ref{ssec:groups}, we identify galaxies in groups using the CiC technique and then separate them into satellites and BGGs.
			It is important to bear in mind that all results shown in this section for BGGs vs. satellites may be affected by the imperfect BGG selection, which will tend to dilute differences between the samples.

\begin{figure}
				\centering
				\resizebox{1\columnwidth}{!}{\includegraphics[width=\columnwidth]{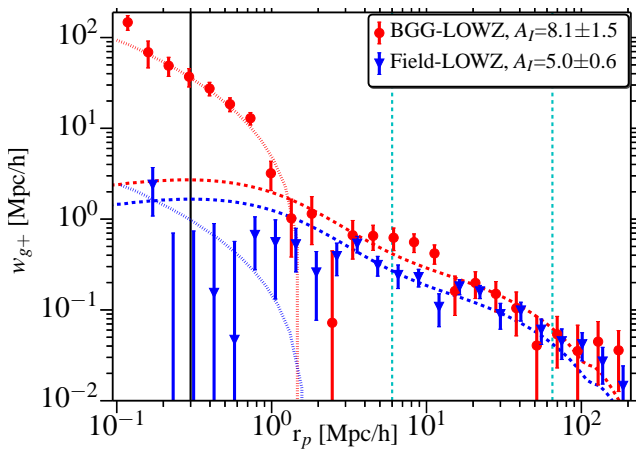} }
				\caption{Density-shape correlation functions for BGGs and field galaxies. BGGs have a much higher intrinsic alignment amplitude (particularly on small scales), which is 
				consistent with BGGs having their shape aligned with their halos and more massive halos having stronger shape alignments.}
				\label{fig:bcg_field_w}
			\end{figure}

\begin{figure}
				\centering
				\resizebox{1\columnwidth}{!}{\includegraphics[width=\columnwidth]{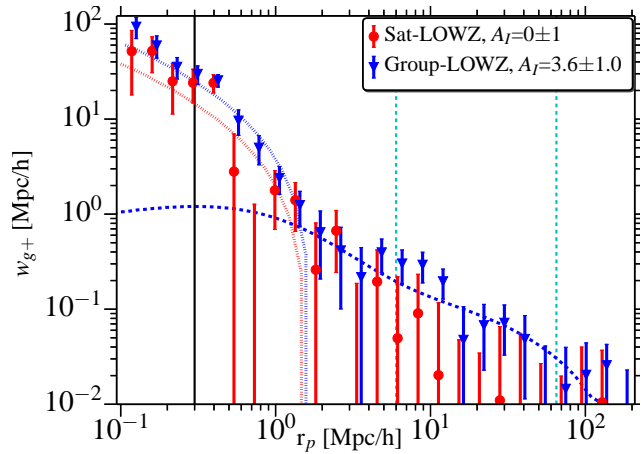} }
				\caption{Density-shape correlation measurements for group galaxies and satellite galaxies correlated with the full LOWZ sample. On large scales, satellite galaxies show no statistically significant shape 
				alignments. On small scales, 
				satellites show some \ia{} signal, which is primarily coming from radial satellite alignments (see also Fig.~\ref{fig:sat_w}).  Group galaxies (satellites + BGGs) exhibit shape alignments with the density field at 
				both small and large scales, with the large scale signal coming primarily from BGGs, and satellites and BGGs both contributing to the small scale signal. }
				\label{fig:group_sat_w}
			\end{figure}

			Figures~\ref{fig:bcg_field_w} and~\ref{fig:group_sat_w} show our measurements for different combinations of galaxies within the groups and the field galaxies. In the plots, the density sample for both signals shown 
			is the same, so the interpretation of large-scale amplitudes includes the same factor of density sample bias. 
			At linear scales, we find that BGGs do tend to align with the 
			tidal field and that BGG alignments are also stronger than that of field galaxies. This effect is consistent with the halo mass dependence of \ia{} seen in Sec.~\ref{ssec:wl_results}, with BGGs living in more massive 
			halos and thus showing stronger alignments.  
			Satellite galaxies on the other hand, do not show any alignments at large scales, with $A_I=0\pm1$. This is discrepant with the luminosity and mass relations of \ia{} observed earlier, but consistent with 
			the expectations from the halo model, where satellites are expected to be radially aligned within the halos and hence will have much weaker signal on large scales.
			Satellites are also the reason that the large-scale \ia{} amplitude for group galaxies is lower than that of BGGs or even field galaxies. 
			
			At small scales, we detect a significant density-shape alignment signal for both satellite galaxies and BGGs. In the BGG-LOWZ correlation, the small scale signal is dominated by BGG-Group 
			correlations (see also Fig.~\ref{fig:bcg_w}). In both cases what we are seeing is the fact that BGGs tend to point preferentially towards their own satellites or, equivalently, that the luminous red satellites of luminous 
			red central galaxies are preferentially located along the major axes of their hosts.  Also, in the satellite-LOWZ correlations, the primary contribution to the small scale signal is from within the groups as well, with 
			satellite-BGG correlations (satellites pointing radially towards their hosts) being the dominant 
			signal (see also Fig.~\ref{fig:sat_w}). This confirms two of the assumptions 
			made in the halo model: satellites do preferentially align radially within the halos,
            and BGGs preferentially align with the shape of their host halo. {The signal
              observed in satellite-satellite correlations can also be explained by the combination
              of radial satellite alignments and the anisotropic distribution of satellites along
              the major axis of the host halo \citep{Faltenbacher2007}.}
			The results shown in Figures~\ref{fig:bcg_field_w} and~\ref{fig:group_sat_w} are likely to be affected by the contamination in our group sample due to our choice of CiC parameters (see discussion 
			in Sec.~\ref{ssec:groups}), but the results in Figures~\ref{fig:sat_w} and~\ref{fig:bcg_w} are more robust since most of the satellites are within $|\Pi|\lesssim10\mpch$ of the BGG (see Fig.~\ref{fig:group_dist}) and the 
			signal for $r_p\lesssim0.6\mpch$ is not likely to be affected by the contamination in the group sample.
			We also note that even though our results do not agree with the null 
			detection of satellite radial alignments in  studies using cluster galaxies \citep[e.g.,][]{Sifon2014,Schneider2013,Hao2011},
			 our sample of LRGs is inherently different. The samples used in the group and cluster studies 
			are dominated by fainter galaxies, which as shown in Sec.~\ref{ssec:result_lum} have lower \ia{} and hence lower the expected signal. Also, those samples are expected to have a small but 
			non-negligible number of late type (blue/disk) galaxies for which \ia{} may arise 
			from angular momentum alignment, which is a second order effect \citep{Hirata2004}. This will further weaken the likelihood of a detection of \ia{} in the cluster studies. Hence, our detection of satellite radial 
			alignments is not necessarily inconsistent with studies using cluster 
			galaxies. Also see Sec.~\ref{ssec:result_comp} for more discussion.
			
			In Fig.~\ref{fig:lowz_many}, we also observe strong environmental dependence of the halo model fitting function amplitude $a_h$. This result could potentially be contaminated by the effects of non-linear 
			bias of the density sample which the halo model does not account for. We factor out this dependence by looking for variations in $a_h$ when the density sample of galaxies is fixed to LOWZ.
			As shown in Fig. \ref{fig:lowz_many}, we fit a power law relation between halo model amplitude $a_h$ and linear galaxy bias, $b_S$,
			\begin{equation}
				a_h=P_h(b_S)^{Q_h}
			\end{equation}
			We find amplitude $P_h=0.007\pm0.002$ and slope $Q_h=4.43\pm0.35$, in the bias range $1.5<b_S<2.6$.
			We note that the power law dependence of small scale IA amplitude ($a_h$) on the large-scale linear bias ($b_S$) is not intuitively obvious. This relation implies that galaxies in more biased and hence more dense 
			regions show higher \ia{} at small scales, even though they are more likely to be effected by processes such galactic mergers and peculiar velocities which can potentially suppress the \ia{} signal. This suggests that 
			effects such as tidal torquing (which counteract mergers, boosting \ia{}) do play an important role in determining \ia{} at small scales. 

		{Also, as described in Appendix \ref{appendix:RSD}, what we measure is the density-weighted intrinsic shear $\widetilde{\gamma_I}=\gamma_I(1+\delta_S)$. On small scales where the $\delta_S$ term cannot be 
		neglected, the 
		non-linear bias of shape sample (which is correlated with the large-scale linear bias $b_S$) will also contribute to produce correlations between $a_h$ and $b_S$. These correlations will likely go as $a_h\propto b_S$ which 
		is weaker than what 
		we observe ($a_h\propto b_S^{4.4}$). Thus, the \ia{} at small scales are still driving the correlations between $a_h$ and $b_S$.}
			Given the small scatter in this relationship between $a_h$ and $b_S$ shown in 
			Fig.~\ref{fig:lowz_many}, it is apparent that for a given sample of density tracers, the large-scale bias of the shape sample is the single best predictor of the small-scale \ia{} amplitude considered in this 
			work, better than luminosity or host halo mass. 
			 Note that the $a_h$ vs. halo mass relation shown in Fig.~\ref{fig:lowz_many} is driven by the BGG-LOWZ data, where the BGG mass can be affected by satellite contamination and mis-centering effects as 
			discussed in 
			Sec.~\ref{ssec:wl_results}.  However, that does not affect our conclusion that the large-scale bias $b_S$ is a more robust predictor of the small-scale amplitude $a_h$ than the galaxy luminosity. 
			These results were all derived with a fixed density tracer sample; we relax this restriction in the following subsection.

		\subsection{Variation of \ia{} with density sample}
			In this section, we address the question of whether alignments of galaxy shapes also depend on the density sample, i.e., do galaxy shapes tend to be more aligned with more massive or biased objects beyond 
			the simple linear dependence on the bias of the density tracer sample. 
			At linear scales, galaxies are biased tracers of large scale tidal fields and thus within the NLA regime we expect this effect to be negligible. Any changes in $w_{g+}$ for the same shape sample and different density 
			samples should be explained by variations in the bias of the density samples, and not by any changes in \ia{} amplitude, $A_I$.  It is unlikely that the same should be true on small scales, where different density 
			samples could have quite different nonlinear bias and environmental effects that will modify the \ia{} beyond what is expected from the large-scale linear bias of the density tracer.
		
\begin{figure}
				\centering
				\resizebox{1\columnwidth}{!}{\includegraphics[width=\columnwidth]{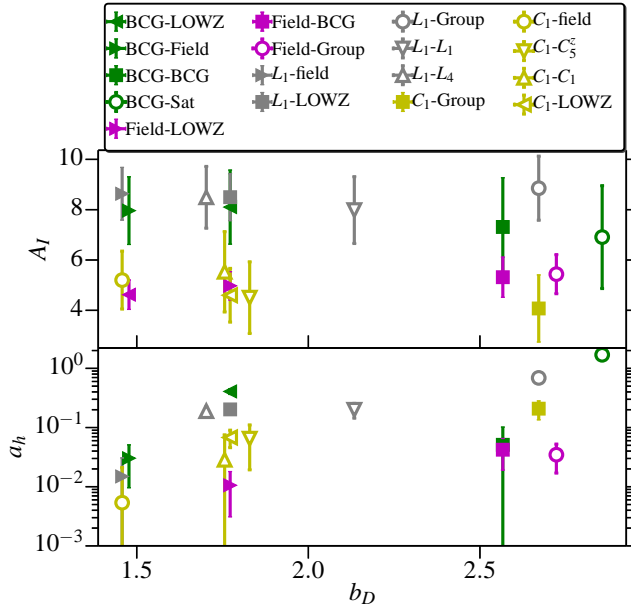} }
				\caption{Variations of \ia{} amplitude with the density samples. Points are color coded according to the shape samples. Consistent with expectations 
				from the NLA model, we do not observe any significant changes in $A_I$ by varying density samples.}
				\label{fig:amp_gamma_density}
			\end{figure}

			The top panel in Fig.~\ref{fig:amp_gamma_density} shows the variations in $A_I$ when a shape sample is correlated with different density samples. For a given shape sample, we do not find any significant 
			changes in $A_I$ with different density samples. The largest 
			change observed is for the field shape sample, where the difference in $A_I$ for field-LOWZ and field-group is $\lesssim 2\sigma$, after accounting for the correlated errors (but see Fig.~\ref{fig:field_theta} for 
			variations observed using a different estimator).
		
\begin{figure}
				\centering
				\resizebox{1\columnwidth}{!}{\includegraphics[width=\columnwidth]{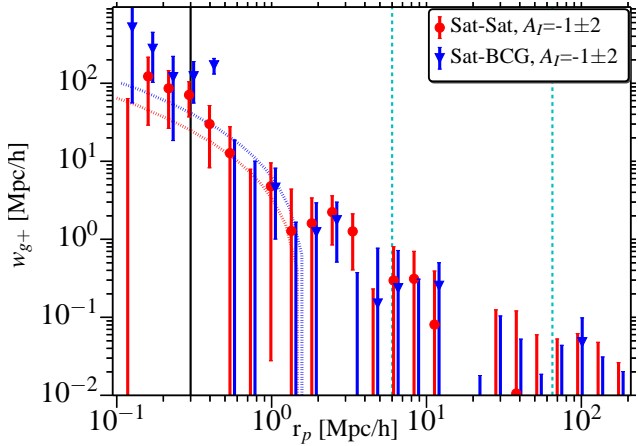} }
				\caption{Variations in the density-shape correlation functions for satellite galaxies correlated with different density samples. 
				}
				\label{fig:sat_w}
			\end{figure}

\begin{figure}
				\centering
				\resizebox{1\columnwidth}{!}{\includegraphics[width=\columnwidth]{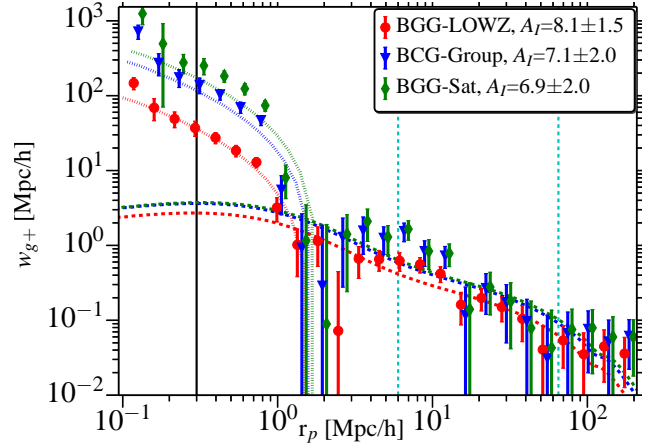} }
				\caption{Variations in the density-shape correlation functions for BGGs correlated with different density samples. For BGG-Group (blue) and BGG-Satellite (green) correlations, halo model does not fit the shape 
				of the correlation function well (blue and green points are horizontally shifted. We donot fit first four bins since they are affected by fiber collisions). Also BGG-Group are lower than BGG-Sat at small scales 
				(they both effectively measure BGG-Sat at small scales) due to higher RR term in the correlation function, since our randoms ($N_R$) are correlated with the density sample ($N_D$) which is higher for group 
				sample due to presence of BGGs which are absent in satellite sample.
 				}
				\label{fig:bcg_w}
			\end{figure}

			As described in Sec.~\ref{ssec:result_env}, on small scales and especially within halos, more complex effects can be important in determining the \ia{} signal for a given shape sample with different density samples. 
			This effect is much harder to interpret using the 1-parameter halo model fitting function that we use in this work. 
			As can be seen in Figs.~\ref{fig:sat_w} and~\ref{fig:bcg_w}, our fitting function does not fully capture the scale dependence of the small-scale 
			signal once the density sample is different from the full LOWZ sample. This is not surprising, since we calibrated the shape of the fitting function to fit signals with LOWZ as the density sample. The scale 
			dependence in the \wgg{} signal suggests that non-linear bias should have significant contributions to the scale dependence of \ia{} signal. This effect, along with some possible variation of \ia{} itself, can be 
			captured if we let more free parameters in the fitting 
			function, but the physical interpretation in that case is not very clear and we defer the study of such effects to future work.
			The bottom panel in Fig.~\ref{fig:amp_gamma_density} also shows the variations in the halo model fit to \wgp{} on small scales. There are significant deviations in $a_h$ observed with variations in the density 
			sample, but as described earlier, it is hard to interpret these in terms of one parameter fitting function; a more detailed physical model is needed.

		\subsection{Putting it all together}\label{ssec:scaling_discussion}
			In this section we briefly discuss and compare the scaling relations of \ia{} amplitudes described earlier and shown in Fig.~\ref{fig:lowz_many}. For NLA amplitude, $A_I$,  we find that $A_I$ scalings with 
			halo mass and luminosity are well described by a power law and scaling with bias $b_S$ is also well described by a linear fit though with larger scatter. All of these scaling are consistent with the observation that 
			more luminous galaxies, living in more massive halos have stronger alignments. It is interesting to note that scaling with luminosity is stronger than scaling with mass, though we caution that mass and luminosity are 
			correlated. To check if this difference in scaling index has any physical meaning, we fit a power law of of the form $M\sim L^{\nu}$ to mass and luminosity of different samples. We get $\nu=1.4\pm0.2$, which 
			compares well with 
			the expectation from simply comparing $A_I$ scalings, $\nu\sim 1.6\pm0.3$. From this we can conclude that these scaling represent same underlying physical effect and one can use either 
			mass or luminosity to predict \ia{} amplitude at large scale. We also note that when fitting for $A_I$ vs $L_r/L_r^p$ scaling, including all the points shown in Fig.~\ref{fig:lowz_many}, the parameters donot change 
			significantly ($\alpha=5.0\pm0.5$ and $\beta=1.33\pm0.25$), which further strengthens our conclusion that either mass or luminosity of sample can be used to predict \ia{} amplitude.

			We also find similar mass and luminosity scaling for the halo model amplitude $a_h$ and we can also draw similar conclusions that both mass and luminosity scaling present same underlying physical effect. It is 
			also interesting to note that mass and luminosity scaling indices for $a_h$ are higher than that for $A_I$, though scatter is also higher.
 			The correlation with linear galaxy bias for $a_h$ is also much stronger, with linear galaxy bias being 
			much better predictor of $a_h$ than mass or luminosity, within the narrow range of our sample. Since the scaling relations with mass/luminosity are steeper for $a_h$ than for $A_I$, this implies that the radial ($r_p
			$) dependence of the intrinsic alignments on $\sim 300h^{-1}$kpc to tens of $h^{-1}$Mpc scales must be changing as the mass/luminosity change. 
			Thus, a single template for intrinsic alignment radial dependence will be insufficient to model intrinsic alignments for future surveys.
			
			As discussed in Sec.~\ref{ssec:results_color} and Sec.~\ref{ssec:results_z}, we do not detect any significant redshift or color evolution of \ia{} amplitudes within the limited range of values probed by our sample (red 
			sequence galaxies in the range $0.16<z<0.36$). However, galaxy properties like mass, 
			luminosity and bias may also depend slightly on the redshift and color. Any such evolution of these properties and \ia{} amplitudes is self-consistently explained by the scaling relations discussed above and we do not 
			detect any additional redshift or color dependence.
			
			To summarize, we find that the best (lowest scatter) predictor of the {\em large-scale} intrinsic alignments signal for a given shape sample is the galaxy luminosity or mass (not its linear bias), whereas the lowest-
			scatter predictor of the {\em small-scale} intrinsic alignments signal for a given shape sample is the linear bias (not its luminosity or mass). However, the shape and amplitude of the small-scale intrinsic alignments 
			signal also depends on the choice of density-tracer sample in a non-trivial way, whereas the large-scale signal only depends in a simple (linear) way on the linear bias of the density sample.
	
		\subsection{Comparison with other works}\label{ssec:result_comp}
			Different studies have used a variety of estimators to measure the \ia{} of galaxies. To compare with their results and also to test for compatibility of different estimators, we present our results using two more 
			estimators, $\langle \theta\rangle$ and $\langle \gamma \rangle$. 	
			
			\begin{equation}
				\langle \gamma \rangle =\frac{S_+D}{SD}
			\end{equation}
			\begin{equation}
				\langle \theta \rangle =\frac{\theta_{SD}}{SD}
			\end{equation}
			$S_+D$ is as defined in Sec.~\ref{ssec:corr}. $\theta_{SD}$ is the angle between the projected major axis of the shape galaxy and the line joining the pair of galaxies.
			 $\langle \gamma \rangle$ measures the average intrinsic shear of the galaxies, while  $\langle \theta\rangle$  measures the tendency of galaxies to align in the direction of other galaxies. In the absence of \ia{},  $
			 \langle \gamma\rangle$ will be consistent with zero while  $\langle \theta\rangle$ will tend to a value of 45$^\circ$. In case of radial alignments,  $\langle \theta\rangle$ will be less than 45$^\circ$ and  $\langle 
			 \gamma \rangle$ will be positive. 
			 
			 Both $\langle \theta\rangle$ and  $\langle \gamma\rangle$ are measured with a single $\Pi$ bin in the range $[-\Pi_{\text{max}},\Pi_{\text{max}}]$. Unless stated otherwise, we take $\Pi_{\text{max}}=100\mpch$. 
			 
\begin{figure}
				\centering
				\resizebox{1\columnwidth}{!}{\includegraphics[width=\columnwidth]{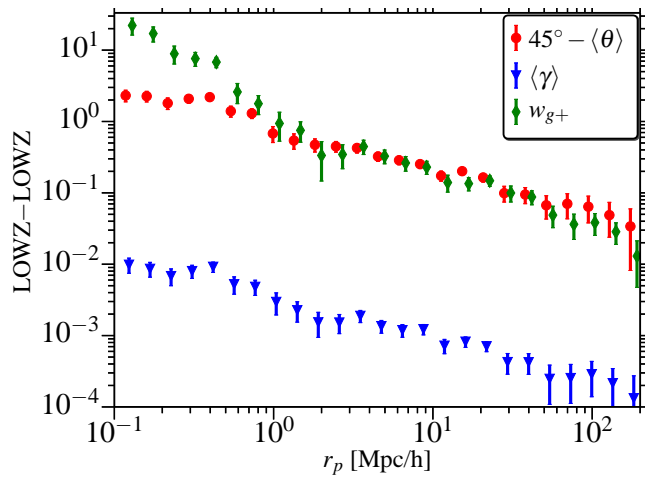}}
				\caption{Comparison of three different estimators for density-shape \ia{}. Note that the similar amplitude of $45^\circ-\langle\theta\rangle$ and \wgp{} has no physical meaning, since they are very different 
				measurements.}
				\label{fig:lowz_theta_gamma}
			\end{figure}

			 Figure \ref{fig:lowz_theta_gamma} shows our measurement of both  $\langle \theta\rangle$  and  $\langle\gamma\rangle$ for the full LOWZ sample. There is good agreement with the three estimators in terms of 
			 scale dependence, with 1-halo to 2-halo transitions around $r_p\sim1-2\mpch$ clearly observable in all three estimators. The flattening of $45^\circ-\langle \theta\rangle$ and $\langle\gamma\rangle$ at small 
			 scales is due to the fact that in taking average, normalization is done  by $SD$ while in $\wgp$ it is done by $RR$. Since at small scales there are more galaxy pairs than would be expected for a random galaxy 
			 distribution, the small scale slope for \wgp{} is 
			 steeper than for $45^\circ-\langle \theta \rangle$ and $\langle\gamma\rangle$. The flattening of the signal at $r_p<0.3\mpch$ could also partly be due to contamination from fiber collision galaxies that we have 
			 included 
			 in our sample, which can potentially bring the signal down if the fiber collision correction is wrong.  Note that similar values of \wgp{} and  $45^\circ- \langle \theta\rangle$ is a coincidence. They represent very 
			 different measurements, and similarities in their magnitudes does not have any physical meaning. 
					
\begin{figure}
				\centering
				\resizebox{1\columnwidth}{!}{\includegraphics[width=\columnwidth]{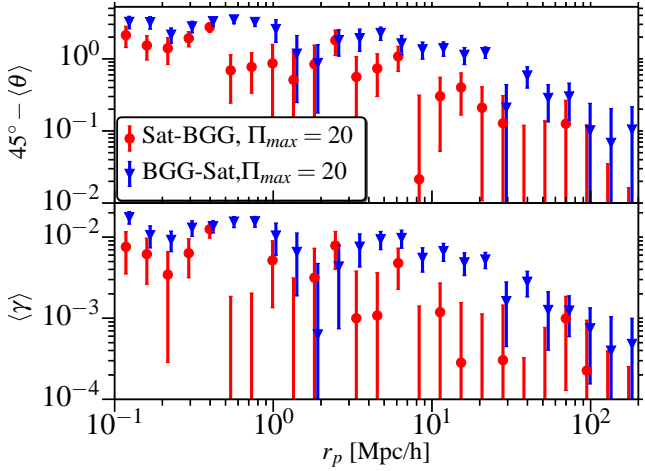}}
				\caption{$45^\circ-\langle \theta\rangle$ and $\langle \gamma\rangle$ for satellite-BGG (red) and BGG-satellite (blue) correlations. $\Pi_\text{max}$ is fixed to $20\mpch$ so that the small-scale signal comes only 
				from within the same group.}
				\label{fig:sat_bcg_theta_gamma}
			\end{figure}

			Fig.~\ref{fig:sat_bcg_theta_gamma} shows the satellite-BGG and BGG-satellite correlations. In this case, we fix $\Pi_{\text{max}}=20\mpch$, so that the signal at small scales, $r_p\lesssim1\mpch$ is only for 
			galaxies 
			within the groups. Our choice of $\Pi$ range here is the same as in Sec.~\ref{ssec:groups}, to select the galaxy pairs within the groups (given that the apparent size of groups along the line of sight is determined by 
			the Fingers of God effect).
			 \cite{Sifon2014} measured the satellite-BGG correlations using spectroscopically selected member for clusters and found a null signal, with $\langle\epsilon_+\rangle=-0.0016\pm0.0020$. For our sample, we find 
			satellite radial shear $\langle\gamma_+\rangle=0.005\pm0.001$, for scales $r_p<1\mpch$. 
			{To provide a plausible explanation for this discrepancy, we assume that the power-law relation between halo model amplitude $a_h$ and $r$-band magnitude $M_r$ in 
			Fig.~\ref{fig:lowz_many} can be extrapolated outside of the luminosity range of the LOWZ sample and can also be extended to $\langle\gamma_+\rangle$ measurement for Sat-BCG type correlations. This assumption, 
			coupled with the fact that galaxies used by \cite{Sifon2014} are 
			on average $\sim 1.5$ magnitude fainter than LOWZ galaxies, implies that the expected signal for \cite{Sifon2014} is lower than our $\gamma_+$ by a factor of $\sim 10-50$ (allowing for 
			some possible variations in luminosity scaling), which is well within the size of their errors. }
			Also, disturbed and late type (blue) galaxies in \cite{Sifon2014} sample will further bring down the expected signal, but their galaxies reside in more massive halos, which could push the signal up again. We 
			 speculate that, with these two effects acting in opposite direction, the expected signal in \cite{Sifon2014} sample will still be lower than observed in our sample.
			
			For $\langle \theta\rangle$, averaging over $r_p<1\mpch$, we find $45^\circ-\langle \theta\rangle=1.5\pm0.2$  for satellite-BGG correlations and $45^\circ-\langle 
			\theta\rangle=3.0\pm0.2$ for BGG-satellite correlations with $\Pi_{\text{max}}=20\mpch$  While satellites and BGGs tend to point towards each other, the effect is stronger for BGGs pointing towards satellites rather 
			than the reverse. Though these results suffer from contamination in our group sample due to our choice of CiC parameters (see discussion in Sec.~\ref{ssec:groups}), the results in Fig.
			\ref{fig:sat_bcg_theta_gamma} as function of $r_p$ are robust since most satellites are within $|\Pi|\lesssim10\mpch$ of the BGGs (see Fig.~\ref{fig:group_dist}).
		
			\begin{figure}
				\centering
				\resizebox{1\columnwidth}{!}{\includegraphics[width=\columnwidth]{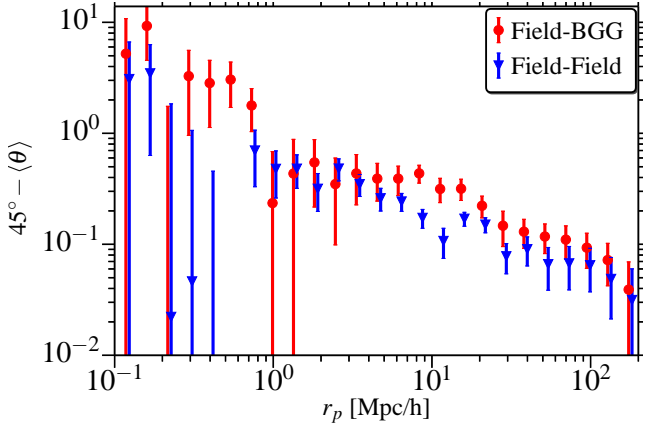}}
				\caption{$45^\circ-\langle \theta\rangle$ for field galaxies with respect to other field galaxies (blue) and BGGs (red). At scales $r_p\sim10\mpch$, field galaxies tend to point more towards BGGs than to other field 
				galaxies, which suggests there is some mass dependent sphere of influence.
				}
				\label{fig:field_theta}
			\end{figure}
		
		Fig.~\ref{fig:field_theta} shows a measurement of $45^\circ-\langle \theta\rangle$ for field galaxies, correlated with other field galaxies (blue) and BGGs (red). $45^\circ-\langle \theta\rangle$ is higher for field-BGG than 
		field-field, which 
		suggests that field galaxies have higher tendency to point towards groups than towards other field galaxies. Averaging over $2\mpch<r_p<20\mpch$, we get $45^\circ-\langle \theta\rangle=0.38\pm0.06$ for 
		field-BGG and $45^\circ-\langle \theta\rangle=0.25\pm0.02$ for field-field correlations (errors are correlated since the shape sample is the same in both cases.).  The deviation from a strict amplitude rescaling, manifested as 
		a change in shape around $10\mpch$, is likely from filamentary structures, with field galaxies residing in filaments around the groups and hence having higher tendency to align with the groups. This result is qualitatively 
		consistent with the findings of \cite{Zhang2013}, who used SDSS DR7 data to reconstruct tidal fields and found that galaxies within filaments tend to have their major axis preferentially aligned with the direction of the 
		filament.
		 More generally, this observation implies that there is a mass dependent sphere of influence within which gravity can 
		align the shapes of galaxies, with BGGs having a larger sphere of influence due to the fact that they reside in higher-mass halos.

			\begin{figure}
				\centering
				\resizebox{1\columnwidth}{!}{\includegraphics[width=\columnwidth]{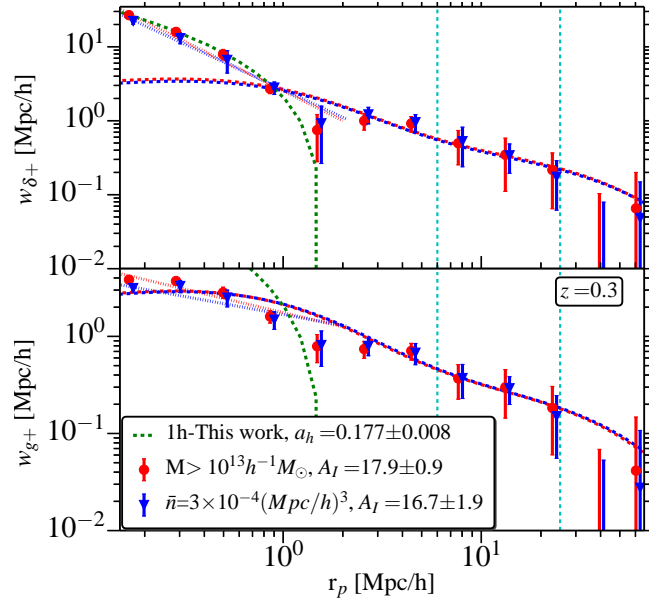}}
				\caption{\wgp{} and $w_{\delta+}$ measurements from the MB-II hydrodynamical simulation \citep{Tenneti2014b}, at redshift $z=0.3$ for a halo mass threshold ($M>10^{13}h^{-1}\Msun$) 
				sample (blue) and a sample defined by a luminosity threshold  
				to have a fixed comoving number density of $\bar{n}=3\times10^{-4}$ (\mpch)$^3$.  
				The dashed green lines show our halo model, while the red and blue dashed lines show the best-fit NLA model.
				}
				\label{fig:tenetti}
			\end{figure}
			
		Fig.~\ref{fig:tenetti} shows results from \cite{Tenneti2014b} for \wgp{} and $w_{\delta+}$ measurements from the MB-II hydrodynamical simulation, at redshift $z=0.3$. We show results from two different samples, the first of 
		which is defined by a halo mass threshold $M>10^{13}h^{-1}\Msun$ (blue) and the second by a luminosity threshold giving a fixed comoving number density of $\bar{n}=3\times10^{-4}$ 
		(\mpch)$^3$. \wgp{} in \cite{Tenneti2014b}  is measured using the shapes of galaxies as defined using the star particles (for those galaxies with $>1000$ star particles), with dark matter subhalos as density tracers 
		($b_D=0.8$), while for $w_{\delta+}$, dark matter particles are the density tracers ($b_D=1$). 
		Both \wgp{} and $w_{\delta+}$ are fit simultaneously at linear scales, assuming they have the same $A_I$ but different $b_D$. On non-linear scales,  \cite{Tenneti2014b} fit a simple power law. The NLA model \ia{} 
		amplitudes are higher for these simulated samples than for LOWZ. The mass threshold sample is comparable to our $L_{1}$ and BGG samples in terms of average halo mass, but its \ia{} are higher than what we observe for 
		these samples by a factor of two. The luminosity threshold sample has a comoving number density similar to the full LOWZ sample, but LOWZ has $A_I$ lower by a factor of $\sim4$. 

		There are several possibilities that could explain the discrepancy between simulations and observations.  The first possibility is some unrealistic aspect of the simulations, including an atypical galaxy shape distribution or a 
		higher degree of alignment between dark matter halo and galaxy shapes than in reality.  
		The second possibility is that this comparison is not completely fair due to the way the galaxy samples were selected in MB-II.  For example, the LOWZ sample has a color cut but the simulated samples do not; in real data 
		we are rejecting some massive galaxies that are even slightly blue and including some lower mass red galaxies, but this effect is not included in the simulated samples.  Also, a mass-selected sample is unrealistic because 
		our mass tracer (luminosity) is known to have a quite high scatter with mass.  Moreover, if this scatter is not present at the same level in the simulated luminosities, that will drastically change the nature of the simulated 
		samples in the direction of having too high average mass and intrinsic alignments.  
		We therefore cannot draw any conclusions from the disagreement in amplitude.  However, it is encouraging that despite the difference in 
		selection, the scaling of $w_{g+}$ with $r_p$ in simulations and the LOWZ observations are comparable.  This suggests that the simulations are likely successful in modeling the basic physical processes that cause the 
		intrinsic 
		alignments, even if the magnitude is not quite correct.
		
	\section{Conclusions}\label{sec:conclusion}
		We have studied \ia{} in SDSS-III BOSS low redshift (LOWZ) galaxies by combining spectroscopic redshifts from BOSS with shape measurements from \cite{Reyes2012}. Using this sample, we have made a robust 
		detection of \ia{} from $0.3$ to $200\mpch$ scales. We find that the NLA model (linear alignment model with non-linear power spectrum) fits the data well for $r_p>6\mpch$, though there are significant 
		deviations at smaller scales. At small scales, we also fitted the data with a halo model-based fitting formula, and find that as long as the density tracer sample is fixed, the \ia{} for different shape samples can be described 
		by simply varying the halo model amplitude, i.e. different shape samples do not introduce any significant scale dependence of their own. However, variations in density samples changes the scale dependence on $
		\lesssim 3\mpch$ scales due to changes in the non-linear bias for different samples.
		
		We also studied the luminosity dependence of \ia{}, finding significant evolution of \ia{} with luminosity, with brighter galaxies having stronger \ia{}. For the density sample comprising of LOWZ galaxies, the luminosity 
		dependence of the \ia{} can be well described by a 
		power law fit in luminosity, with power law slope of $\beta=1.30\pm0.27$ for $A_I$ (NLA model amplitude) and $\beta_h=2.1\pm0.4$ for the small-scale halo model fitting function amplitude. While extrapolation beyond 
		the range of luminosities explored here must be done cautiously, this luminosity dependence likely has important implications for future surveys like LSST and 
		Euclid, which will go to much fainter magnitudes than SDSS, and hence  overall \ia{} contamination for shear studies in such surveys should go down. This, however, does not mean that \ia{} contamination will not be
		important as the statistical error budget for these surveys is also far smaller than for SDSS.

		We calculated the average halo mass of galaxies in various samples using weak lensing and then studied the mass dependence of \ia{}. We find that the \ia{} amplitude (on both small and large scales) increases with halo 
		mass, and the mass-\ia{} amplitude relations are well described by a power law in mass. This result is consistent with luminosity evolution as well and taken together, both luminosity and mass evolutions imply that 
		brighter and more massive galaxies show stronger \ia{}. 
		
		We do not find any significant redshift dependence of \ia{}  within the limited range of $0.16<z<0.36$. This is consistent with the linear alignment model, which assumes 
		that \ia{} are imprinted at earlier times during the formation of galaxy. Other effects at later times can in principle introduce some redshift dependence, but the effect seems to be too small to be detected in our narrow 
		redshift range.
		
		We also do not find any significant evolution of \ia{} with color, within the red sequence. We split the sample into 5 color bins with each bin containing 20 per cent of the LOWZ sample. There is some luminosity evolution 
		within our color samples, and any \ia{} variation that we see within these samples can be explained by the luminosity scalings.
		
		We identified the galaxies in groups using the CiC technique as described in Sec.~\ref{ssec:groups} and studied the environment dependence of \ia{}. We find large scale alignments for BGGs in our sample as well as 
		strong alignments of BGGs shapes with satellites. Our results are consistent with the halo model as well as observations from simulations where BGGs are found to be aligned with their host dark matter halos (as traced 
		by satellites), which in turn have large scale alignments as described by linear alignment model. {We also detect small scale alignments of satellite shapes, consistent with radial alignments assumed in halo model 
		and observed in simulations (this conclusion is based on the assumption that BGGs are the center of halos). 
		We reiterate that  the CiC parameters used in this work were optimized for the higher mass SDSS LRG sample, not the LOWZ sample. This could potentially lead to some 
		contamination in our group sample. However, based on our estimates of the importance of this effect, it does not significantly affect our overall conclusions about the environmental dependence of intrinsic alignments.}

		Finally, comparison with previous studies using a variety of estimators suggests that our results are either in good agreement or the differences can be explained by differences in 
		the sample selection, most notably the average luminosity of the galaxies.

		To summarize, our results suggest that mass and luminosity are the most important factors in determining the \ia{} of red galaxies, with color (within the red sequence) and redshift effects being subdominant. The best 
		predictor (least scatter) for \ia{} at large scales is 
		the halo mass/luminosity of the galaxies, while at small scales, the (large-scale) linear bias is the best predictor assuming that the density tracer sample is kept fixed. Comparing our halo fitting function and NLA amplitudes to those observed in simulations should help in better 
		understanding 
		of processes involved in galaxy formation and how baryonic matter and dark matter influence each other within the halos.	Moreover, our results and constraints on the NLA model and the small-scale fitting function should 
		also be useful in putting priors on models of \ia{} used in mitigation schemes \citep[see for example,][]{2005A&A...441...47K,Bridle2007,2010A&A...523A...1J} for upcoming large surveys that aim to use weak 
		lensing to measure the equation of state of dark energy to extremely high precision.

	\section{Acknowledgements}
This work was supported by the National Science Foundation under
Grant. No. AST-1313169. RM was also supported by an Alfred P. Sloan Fellowship. SM is supported by World Premier International Research Center
Initiative (WPI Initiative), MEXT, Japan, by the FIRST program ``Subaru
Measurements of Images and Redshifts (SuMIRe)'', CSTP, Japan. We thank Benjamin Joachimi, Alina Kiessling,
Marcello Cacciato, Jonathan Blazek, and Cristobal Sif{\'o}n for useful discussions about
this work. {We thank the anonymous referee for useful feedback that improved the paper.}
	
Funding for SDSS-III has been provided by the Alfred P. Sloan Foundation, the Participating Institutions, the National Science Foundation, and the U.S. Department of Energy Office of Science. The SDSS-III web site is http://www.sdss3.org/.

SDSS-III is managed by the Astrophysical Research Consortium for the Participating Institutions of the SDSS-III Collaboration including the University of Arizona, the Brazilian Participation Group, Brookhaven National Laboratory, Carnegie Mellon University, University of Florida, the French Participation Group, the German Participation Group, Harvard University, the Instituto de Astrofisica de Canarias, the Michigan State/Notre Dame/JINA Participation Group, Johns Hopkins University, Lawrence Berkeley National Laboratory, Max Planck Institute for Astrophysics, Max Planck Institute for Extraterrestrial Physics, New Mexico State University, New York University, Ohio State University, Pennsylvania State University, University of Portsmouth, Princeton University, the Spanish Participation Group, University of Tokyo, University of Utah, Vanderbilt University, University of Virginia, University of Washington, and Yale University.

	\bibliographystyle{mn2e2}
%		\sukhdeep{Using an edited file for mnras bib. File from {http://user.astro.columbia.edu/~williams/mn2ebst/}. More edits to get arxiv citations right} 

	\bibliography{sukhdeep_paper_final2}
	
	\appendix
\onecolumn
\section{Effect of Redshift space distortions on intrinsic alignments}\label{appendix:RSD}

{The effect of redshift space distortions on the galaxy clustering
signal was first studied by \citet{Kaiser:1987}.} In this appendix we show the effect of redshift space distortions on the
intrinsic alignment correlation functions. For simplicity, we will consider
configuration space, assume that the plane parallel approximation holds, and
that the $z$ direction corresponds to the line-of-sight. The following
coordinate transformation maps real space to redshift space,
\begin{equation}
        x' = x; \,\,y' = y; \,\,z' = z + v \mu; \,\,,
\end{equation}
where the primed quantities correspond to redshift space, we have assumed units
with $H_0=1$, $v$ is the magnitude of the velocity at the real space position
(positive value corresponds to motion away from the observer) and $\mu$ is the
cosine of the angle between the velocity and the line-of-sight.

Let us begin by considering the quantity $\greal$ which is the intrinsic
alignment of a shape sample of galaxies with respect to matter in real space.
Since the intrinsic shear is measured at the position of the shape sample of
galaxies, in practice, we measure the density-weighted intrinsic shear,
$\tgreal$, where
\begin{align}
        \tgreal&=\greal(1+\delta_\rmS^{\rm real})\,\nonumber\\
        &=\greal(1+b_\rmS\delta^{\rm lin})\,\nonumber \\
        &=\greal+{\cal O}(\delta^{\rm lin}\delta^{\rm lin})\,,
\end{align}
where $\delta_\rmS^{\rm real}$ is the overdensity of the shape sample of
galaxies and $b_\rmS$ its bias with respect to the matter density fluctuation
$\delta^{\rm lin}$. The last equality follows from the proportionality $\greal
\propto \phi_{\rm p}$ (see Equation~\ref{eqn:gamma_phi}), and
$\phi_p\propto\delta^{\rm lin}$ from the Poisson equation. Here $\delta^{\rm
lin}$ is the matter density fluctuation. Thus if we restrict to linear order (as
we will throughout this appendix), there is no correction to the intrinsic shear
due to the density weighted measurement. Therefore, we will not make a
distinction between the intrinsic shear and its density weighted counterpart,
hereafter.

In this appendix we will work out the relation between $\greal$ and the
corresponding redshift space quantity, $\gred$. Since the shear distortion comes
from the tidal field at the true position of the galaxy, and is unchanged by redshift-space distortions, the only effect of RSD is to change the apparent position of the galaxy.  This will result in its contributing to the 3D correlation function $\xi_{g+}$ or $\xi_{++}$ at a different value of $\Pi$, but without changing the intrinsic shear value itself.  $\gred$ and $\greal$
are therefore related by
\begin{equation}
\gred(x',y',z')=\greal(x,y,z)\,.
\end{equation}
The Fourier transform of the left hand side gives
\begin{align}
        \gred(k'_x,k'_y,k'_z)&= \int dx' dy' dz' \gred(x',y',z') e^{i(k'_x x'+k'_y y'+k'_z z' )}\,, \nonumber\\
				      &= \int dx'\,dy'\,dz'\,\greal(x',y',z'-v\mu) e^{i(k'_x x'+k'_y y'+k'_z z' )} \nonumber\\
				      &= \int dx\,dy\,dz\, \greal(x,y,z) e^{i(k'_x x+k'_y y+k'_z [z+ \mu v] )}\frac{dz'}{dz} 
\end{align}
where $dz'/dz=1+ \mu\partial_z v$. We can combine this term and expand the
exponential as a Taylor series to obtain
\begin{align}
        \frac{dz'}{dz}e^{ik'_z\mu v} &= (1+ \mu\partial_z v) \sum_{n=0}^{\infty} \frac{ i^n (k'_z\mu v)^n}{n!} \nonumber\\
					        &= 1 + \mu\partial_z v + i k'_z \mu v  + {\cal O}(\delta^{\rm lin}\delta^{\rm lin}) \nonumber \\
					        &= 1 - 2 \mu^2 f \delta^{\rm lin} + {\cal O}(\delta^{\rm lin}\delta^{\rm lin})
\end{align}
Here we have used $\mu\partial_z v= i k'_z \mu v=-\mu^2\dot\delta^{\rm
lin}=-\mu^2f\delta^{\rm lin}$, where the last two equalities are valid at linear
order. Therefore,
\begin{align}
        \gred(k'_x,k'_y,k'_z) &= \int dx\,dy\,dz\greal(x,y,z) e^{i(k'_x x'+k'_yy+k'_z z )} [1+{\cal O}(\delta^{\rm lin})] \nonumber \\
					&= \greal(k'_x,k'_y,k'_z) + {\cal O} (\delta^{\rm lin}\delta^{\rm lin})
\end{align}
This implies that there is no linear order correction to the intrinsic shear
field while going from real to redshift space.

Now consider a density sample characterized by its overdensity $\delta_\rmD$. Number density
conservation relates the redshift space overdensity to the real space
overdensity such that
\begin{equation}
        (1+\delta^{\rm red}_\rmD)dx'dy'dz'= (1+\delta^{\rm real}_\rmD) dx\,dy\,dz\,,
\end{equation}
Therefore,
\begin{align}
        \delta^{\rm red}_\rmD&= (1+\delta^{\rm real}_\rmD) \left[\frac{dz'}{dz}\right]^{-1} - 1\nonumber\\
               &= (1+\delta^{\rm real}_\rmD) \left[ 1 -  \mu\partial_z v \right] - 1 + {\cal O}(\delta\delta)\nonumber\\
               &= \delta^{\rm real}_\rmD - \mu\partial_z v + {\cal O}(\delta\delta)\nonumber\\
               &= \delta^{\rm real}_\rmD + f \mu^2 \delta^{\rm lin} + {\cal O}(\delta\delta)\nonumber\\
               &= \delta^{\rm real}_\rmD\left(1+\frac{f}{b_\rmD}\mu^2\right)+ {\cal O}(\delta\delta)
\end{align}
The linear order correction of $(1+f/b_\rmD\mu^2)$ when going from real
coordinates to redshift coordinates is valid both in configuration and
Fourier space.

The auto-power spectrum of the intrinsic shear and its cross power spectrum with
the galaxy overdensity field in redshift space and real space are therefore
related by 
\begin{align}
        P^{\rm red}_{++}(\vc{k}) &= \langle  \widetilde{\gamma}_I^{\rm red}(\vc{k})  \widetilde{\gamma}_I^{\rm red}(\vc{k}) \rangle \nonumber\\ 
        &= \langle\greal(\vc{k}) \greal(\vc{k}) \rangle + {\cal O}(\delta^3) \nonumber\\ 
        &=P^{\rm real}_{++}(\vc{k}) + {\cal O}(\delta^3) \\
        P^{\rm red}_{g+}(\vc{k}) &= \langle \delta_\rmD^{\rm red}(\vc{k})  \widetilde{\gamma}_I^{\rm red}(\vc{k}) \rangle \nonumber\\
        &= \left(1+\frac{f}{b_\rmD}\mu^2\right)\langle \delta_\rmD^{\rm real}(\vc{k})\greal(\vc{k}) \rangle + {\cal O}(\delta^3) \nonumber \\
        &=\left(1+\frac{f}{b_\rmD}\mu^2\right)P^{\rm real}_{g+}(\vc{k}) + {\cal O}(\delta^3) \,,
\end{align}
while the galaxy-galaxy power spectrum in real and redshift space are
related by 
\begin{align}
        P^{\rm red}_{gg}(\vc{k}) &= \langle \delta_\rmD^{\rm red}(\vc{k})\delta_{\rm S}^{\rm red}(\vc{k})\rangle \nonumber\\
        &= \left(1+\frac{f}{b_\rmD}\mu^2\right)\left(1+\frac{f}{b_{\rm S}}\mu^2\right)\langle \delta_\rmD^{\rm real}(\vc{k})\delta_{\rm S}^{\rm real}(\vc{k}) \rangle + {\cal O}(\delta^3)\nonumber\\
        & =\left(1+\frac{f}{b_\rmD}\mu^2\right)\left(1+\frac{f}{b_{\rm S}}\mu^2\right)P^{\rm real}_{gg}(\vc{k}) + {\cal O}(\delta^3)\,.
\end{align}
		
\end{document}